# On-Demand Transit User Preference Analysis using Hybrid Choice Models

**Nael Alsaleh**
Laboratory of Innovations in Transportation (LiTrans)
Department of Civil Engineering,
Toronto Metropolitan University (Formerly Ryerson University), Toronto, ON Canada, M5B 2K3
Email: nael.alsaleh@ryerson.ca

**Bilal Farooq (**Corresponding author**)**
Laboratory of Innovations in Transportation (LiTrans)
Department of Civil Engineering,
Toronto Metropolitan University (Formerly Ryerson University), Toronto, ON Canada, M5B 2K3
Email: bilal.farooq@ryerson.ca

**Yixue Zhang**
Department of Geography & Planning,
University of Toronto, Toronto, ON Canada, M5S 1A1
Email: yixue.zhang@mail.utoronto.ca

**Steven Farber**
Department of Geography & Planning,
University of Toronto, Toronto, ON Canada, M5S 1A1
Email: steven.farber@utoronto.ca

**Abstract**

In light of the increasing interest to transform the fixed-route public transit (FRT) services into on-demand transit (ODT) services, there exists a strong need for a comprehensive evaluation of the effects of this shift on the users. Such an analysis can help municipalities and service providers to design and operate more convenient, attractive, and sustainable transit solutions. To understand the user preferences, we developed three hybrid choice models: integrated choice and latent variable (ICLV), latent class (LC), and latent class integrated choice and latent variable (LC-ICLV) models. We used these models to analyze the public transit user's preferences in Belleville, Ontario, Canada. Hybrid choice models were estimated using a rich dataset that combined the actual level of service attributes obtained from Belleville's ODT service and self-reported usage behaviour obtained from a revealed preference survey of the ODT users. The latent class models divided the users into two groups with different travel behaviour and preferences. The results showed that the captive user's preference for ODT service was significantly affected by the number of unassigned trips, in-vehicle time, and main travel mode before the ODT service started. On the other hand, the non-captive user's service preference was significantly affected by the Time Sensitivity and the Online Service Satisfaction latent variables, as well as the performance of the ODT service and trip purpose. This study attaches importance to improving the reliability and performance of the ODT services and outlines directions for reducing operational costs by updating the required fleet size and assigning more vehicles for work-related trips.

## 1. Introduction

In low demand areas, on-demand public transit (ODT) service is emerging as an attractive solution that can address key issues associated with the existing fixed-route public transit (FRT) service. In such settings, as a compromise between the occupancy of the vehicle and their operational cost, FRT is often operated with very low frequency, limited operating hours, and inadequate spatial coverage (Papanikolaou et al., 2017; Sanaullah et al., 2021). These circumstances can make FRT inconvenient and unattractive for residents, especially for non-captive users, who have alternative travel options available to satisfy their travel needs. Alternatively, the main philosophy behind an ODT service is to minimize the operational cost of the service by dynamically adjusting to the user schedules instead of reducing the service frequency. The ODT proponents argue that it may enhance the service quality as well as reduce the operational cost and greenhouse gas emissions in low-demand and low-density areas compared with the FRT service (Papanikolaou et al., 2017; Diana et al., 2007). From publicly available data from 2018 to 2020 (Mobility Innovation Lab, 2022; American Public Transportation Association, 2022), it is clear that transit operators in North America are increasingly interested in providing ODT services in low density settings (see Figure 1). Among the existing ODT projects in North America, a few examples include Belleville in Ontario, Cochrane in Alberta, Regina in Saskatchewan, Winnipeg in Manitoba, Bowen Island in British Columbia, Montreal in Quebec, Gwinnett County in Georgia, Austin in Texas, Antioch in California, Grand Rapids in Michigan, Eugene in Oregon, Albany in New York, St. Louis in Missouri, Jacksonville in Florida, Tucson in Arizona, Columbus in Ohio, and King County in Washington (Mobility Innovation Lab, 2022; American Public Transportation Association, 2022). Unlike ride-hailing, in the ODT service the fare is predominantly fixed, fleet-size remains stable, vehicles are mainly owned by an organization (e.g., municipality), and drivers receive a fixed salary.

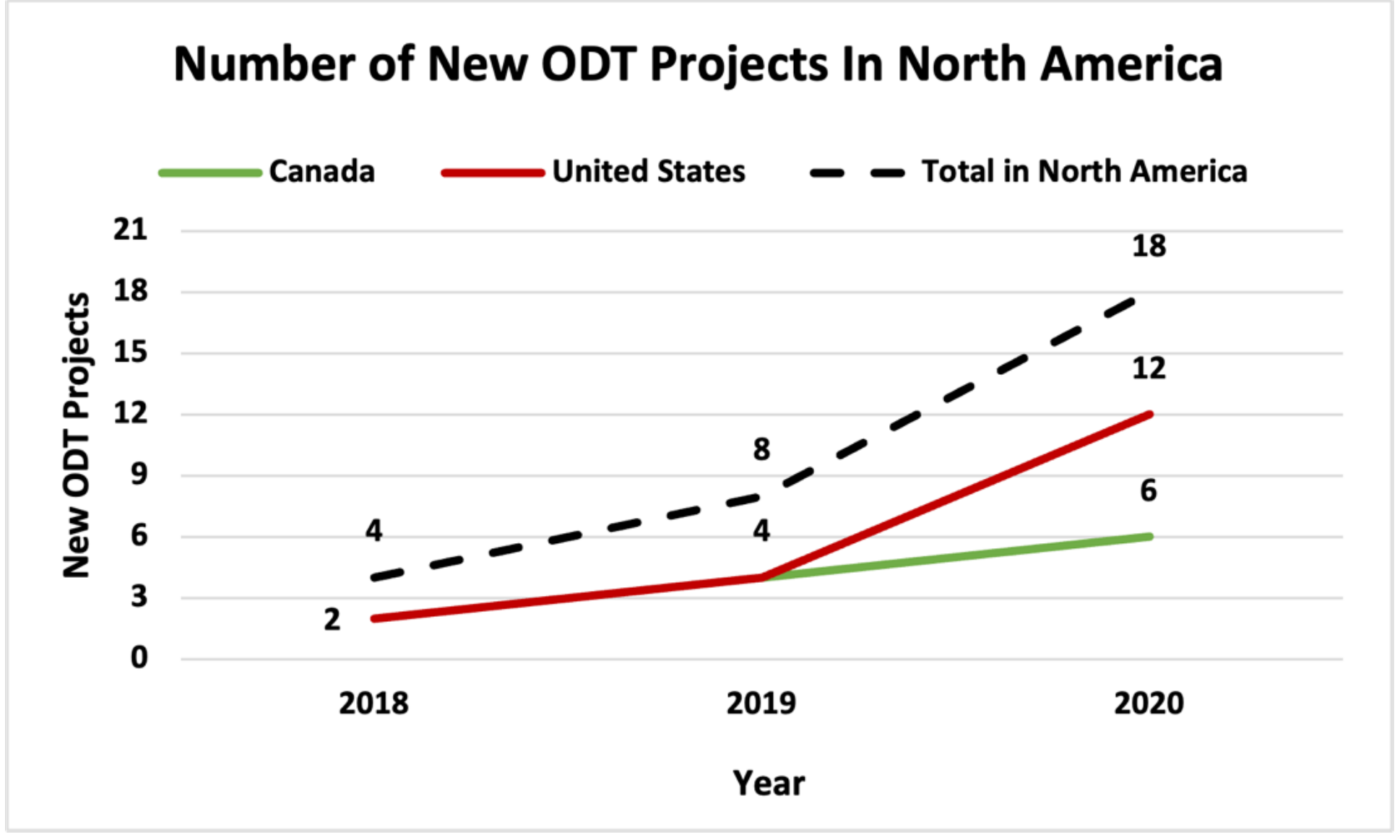

**Figure 1.** Number of new ODT projects in North America per year.

In September 2018, one of the first ODT projects started in Belleville, Ontario, Canada, where the city collaborated with a private sector partner to convert its late-night FRT service operated on route 11 (RT-11) into an ODT service. The resulting service altered its operating routes according to the real-time spatio-temporal demand the system received. The service was no longer limited to the original RT-11 stops. Instead, it covered the entire city based on the pickup and dropoff requests. Users booked their trips online using the service app, website, or by calling the phone number. The Belleville ODT service runs for 5 hours during the weekends and 3 hours during the weekdays, from 7:00 PM to 12:00 AM and from 9:00 PM to 12:00 AM, respectively (Alsaleh and Farooq, 2021; Sanaullah et al., 2021).

Certain key features of the ODT service, for instance, route flexibility and larger service areas, may render the service inconvenient for some users. Waiting time of the user (the difference between the actual pickup time and the requested pickup time) and in-vehicle time might be different each time an ODT trip request is placed, even if the requests are made at the same time of the day or day of the week, from the same origin to the same destination. Unlike the FRT, which runs along a predefined fixed route with a predefined schedule, an ODT user's waiting and in-vehicle times would depend on the number of detours the bus makes. Moreover, the performance of an ODT service is affected by the number of requests the system receives, especially when the supply side (maximum number of available vehicles) is limited. Therefore, users might experience long waiting and in-vehicle times when the ODT service encounters unusually high demand, for instance, during public holidays, special events, and on frigid winter days. This pattern was confirmed by the study conducted by Sanaullah et al. (2021), which showed that both the number of ODT trips and service performance were unstable during the first three months of operation, starting October 2018. The average waiting time in November and December 2018 was 7 minutes (30%) longer than the average waiting time before and after, due to the holiday season related increase in the number of requested trips. The preference of ODT as a sustainable mode choice for users may vary based on their experience and attitudes towards the service. Thus, for the sustainable adoption of the ODT service, this study investigates the following key questions:

- *Do the attitudes and perceptions towards the ODT service affect user's preference between the FRT and ODT services?*
- *Is there a difference in the service preference between the ODT captive and non-captive users?*
- *What can transit operators do to attract more users and influence the non-captive users to become regular ODT service users?*

To this end, we developed three hybrid choice models to explain the ODT user's service preference between the FRT and ODT services. These models include an integrated choice and latent variable (ICLV) model, latent class (LC) model, and latent class integrated choice and latent variable (LC-ICLV) model. We used these models in our study to alleviate the limitations of traditional discrete choice models by accounting for the subjectivity of human behaviour and/or the preference heterogeneity among the latent classes of population in the decision-making process. For comparison purposes, a multinomial logit (MNL) model was used as a base model. Based on our findings, we further provide suggestions and practical implications to help the municipalities and operators design and operate more convenient and attractive ODT services. The data used in the modelling process are fused from the actual level of service attributes obtained from Belleville's ODT service and self-reported data obtained from a revealed preference survey of Belleville's ODT users. To the best of our knowledge, no previous work in the literature exists that has developed advanced behavioural models for the ODT users to understand their preferences and based on which has provided detailed design and operational recommendations.

The remainder of the paper is organized as follows. Section 2 discusses the recent literature on the user preferences for flexible transit services and discrete choice modelling techniques. Section 3 describes the dataset, followed by Section 4, which explains the framework and the specifications of the MNL, LC, ICLV, and LC-ICLV models. Section 5 presents and illustrates the estimation results of the developed models. We provide a detailed discussion with design and operational recommendations in Section 6 and a conclusion in Section 7.

## 2. Background

This section reviews previous studies that modelled and analyzed user preference for flexible transit services. We also present recent advancements in the use of discrete choice modelling to address travel behaviour issues. It is important to note that currently, there is a lack of literature that directly models the behaviour of the ODT users in general and their service preferences in particular. This lack is due to the limited number of ongoing ODT projects in the market and availability of revealed as well as stated preference data on the topic.

2.1 User Preference for Flexible Transit Services

The conventional fixed-route public transit (FRT) service offers a set of well-defined pickup/dropoff stops and times to serve passengers along a predefined route (Hall et al., 2018). However, the availability of smartphones and network connectivity has enabled various flexible forms of public transit services such as demand-responsive transit (DRT), paratransit, and mobility on demand (MOD) transit services (Ma et al., 2020). DRT system has two main operational types: microtransit and ODT services (Klumpenhouwer, 2020). Both services offer flexible transit to the public, but they differ in terms of vehicle size and purpose. Microtransit is using small-sized vehicles to complement the FRT service in low demand and low-density areas, where the FRT service is limited or not provided (Jain et al., 2017; Ma et al., 2020). ODT service is also used in low demand and low-density areas, but usually as a substitute for the FRT service using the existing infrastructure (buses and stops) and fare system (Alsaleh and Farooq, 2021). Paratransit provides a door-to-door service for impaired passengers, for instance seniors and people with disabilities, within a predefined area (Torkjazi and Huynh, 2019). MOD transit is an integrated transit system that operates FRT services along the major roads and ride-hailing and/or microtransit services in the low-demand areas (Yan et al., 2019a; Zhao et al., 2019).

Understanding user's choice behaviour towards flexible transit services can help transit agencies improve the service quality and increase their market uptake. Nevertheless, only a few studies have investigated the user preference for DRT services. Anspacher et al. (2004) developed an ordered logit model to examine user preference for a proposed microtransit service as an access mode to and from the rail transit station in San Francisco Bay Area, California. The estimation results revealed that park and ride users in the suburban communities, as well as travellers who carpooled or used public transit to get to and from the rail transit station, were more willing to use the proposed microtransit service. Moreover, travellers' willingness to use the proposed microtransit service increased as the distance to the nearest transit station increased. Furthermore, elderly and travellers who had more private vehicles were less willing to use the proposed service. Tarigan et al. (2010) conducted a revealed preference survey on the microtransit users in Bandung, Indonesia, to investigate the impact of negative experiences on user's willingness to use the service using an ordered probit model. The study found that the cost, practicality, and the built environment characteristics had the highest impact on men's willingness to use the microtransit service, whereas the service safety influenced the women most.

Khattak and Yim (2004) conducted a stated preference survey in the San Francisco Bay Area, California, to examine user's willingness to use a hypothetical personalized DRT service. The survey results showed that most of the respondents were willing to use the proposed service and pay more than the FRT service. For most respondents, the reliability, cost, and pickup and dropoff locations were the most important attributes of the proposed service. Yu et al. (2017) also investigated user's willingness to use a hypothetical ODT service in Jinan Qilu Software Park, China. The survey results revealed that women, enterprise employees, participants with a college degree or higher, and participants with higher income were more willing to use the proposed service. Zhang et al. (2021) conducted a web-based revealed preference survey on the ODT riders in Belleville, Ontario, to examine their satisfaction with the service and capture the relationship between satisfaction and activity participation. The results revealed that the participants were most satisfied with driver's qualifications and attitudes. However, they were least satisfied with the waiting time and reliability of the service. Moreover, the study found that the user's willingness to engage in activities was most affected by their satisfaction with the service reliability and performance.

Paratransit is an important service for the elderly and passengers with disabilities to meet their mobility needs. However, it is more expensive to operate in comparison to the conventional FRT service (Kaufman et al., 2016). Thus, several studies have been conducted on identifying the main impediments for the elderly and/or passengers with disabilities to use microtransit service. For instance, Miah et al. (2020) conducted an interview-based survey with 128 paratransit users in Arlington, Texas, to capture the main barriers for impaired passengers to adopt microtransit. The study reported that the lack of spatial coverage and accessibility, as well as the difficulty to use, were the main impediments for paratransit users to use microtransit services. Jittrapirom et al. (2019) found that the main reasons for the elderly not using the microtransit service in the Netherlands were the availability of more comfortable travel options, inconvenience of the service, and unfamiliarity with the service app. Leistner and Steiner (2017) also found that unfamiliarity with smartphones and apps was the main barrier for the elderly to use app-based transportation services.

Some studies have conducted stated preference surveys to investigate user's willingness to use MOD transit services. For instance, Yan et al. (2019a) developed a mixed logit model to examine user response to a proposed MOD transit service (integration of ride-hailing and public transit services) at the University of Michigan Ann Arbor campus. The results showed that replacing the FRT service in the low demand areas with ride-hailing services can slightly increase the ridership of the public transit system, while minimizing the operational cost. Yan et al. (2019b) used an ordered logit model to investigate disadvantaged user's (disabled, elderly, low-income, and carless travellers) preferences for a proposed MOD transit service (integration of microtransit and FRT services) versus the FRT service in Detroit and Ypsilanti, Michigan. The estimation results indicated that males, highly educated people, users who have not heard or used the ride-hailing services before, and users who had a negative experience with the FRT service, were more likely to prefer the proposed system. Saxena et al. (2020) conducted a stated preference survey in the Northern Beaches area of Sydney, Australia, to examine user preferences for a proposed MOD transit service that combined a microtransit service with an FRT service. The study probabilistically segmented the participants into two latent groups based on their mode choice behaviour. The estimation results of the class membership functions revealed that individuals with work-based trips were most likely to belong to the first group, whereas individuals taking non-work trips had a higher tendency to belong to the second group. The estimation results of the choice model showed that individuals in the first group had a higher uptake for the proposed service than those in the second group. Moreover, the attributes of the proposed service, including in-vehicle time, travel cost, and access time, were found to have a significantly negative impact on the individuals belonging to the first group. In contrast, the individuals belonging to the second group were found to be indifferent towards these attributes.

## 2.2 Discrete Choice Modelling

Traditional random utility maximization (RUM) based discrete choice models have been used extensively in modelling the individual travel behaviour (Ben-Akiva and Lerman, 1985; Pryanishnikov and Zigova, 2003). These models are generally based only on the observable variables and do not consider the impact of decision-makers' attitudes, perceptions, and lifestyles on their decision-making behaviour (Ababio-Donkor et al., 2020). Furthermore, the parameter estimates of these models might be ambiguous as a consequence of assuming that the entire population has the same preference structure and choice behaviours, given that the transportation market has different segments (Hurtubia et al., 2014; Alizadeh et al., 2019; Zhou et al., 2020). Therefore, several RUM based hybrid choice models have been developed to overcome these limitations, for instance, integrated choice and latent variable (ICLV) model, latent class (LC) model, and the latent class integrated choice and latent variable (LC-ICLV) model.

The ICLV model, which was proposed by McFadden (1986), is a modified version of RUM based logit model. It incorporates the individual's attitudes, perceptions, norms, or lifestyles into the systematic component of the utility function as latent variables (Alizadeh et al., 2019; Ababio-Donkor et al., 2020).

Several studies have incorporated latent constructs into traditional choice models in order to improve their explanatory power when modelling individual travel behaviour. To name only a few, Cantillo et al. (2015) used ICLV model to capture the impact of the attractiveness and the safety/security latent variables in addition to other directly measured variables on the pedestrian crossing behaviour in urban roads. The study reported that both latent variables had a significant impact on pedestrians crossing behaviour. In another study, Belgiawan et al. (2016) estimated an ICLV model to explain the car purchase decision among the Indonesian students. The results showed that the arrogance and prestige latent variables in addition to some sociodemographic variables had a significant impact on the car purchase decisions.

The LC model is another extension of logit model, which assumes that the population can be probabilistically divided into a discrete number of latent classes, such that the perceptions or preferences are homogenous within the same class and heterogeneous across the different classes (Boxall and Adamowicz, 2002; Walker and Ben-Akiva, 2002; Hurtubia et al., 2014; Alizadeh et al., 2019). The LC framework has been used in several studies to understand the travel behaviour of different market segments as well as identify their needs. For instance, Zhou et al. (2020) used the LC model to explain the travel mode and airline choice behaviour of two types of travellers. The first type represented business travellers with higher incomes and the second type represented personal travellers with lower incomes. The results indicated that the choice behaviour of the two types of travellers was significantly different. It is worth noting that a mixed logit (ML) model can also account for the taste variations of decision-makers by allowing its parameters to vary with a known population distribution across decision-makers. However, such a model requires specifying the distribution of the parameters before-hand (Shen, 2009). The LC was compared to MNL, ML, or both models in the previous literature, and the results showed that the LC model was superior (Massiani et al., 2007; Hess et al., 2009; Shen, 2009; Greene and Hensher, 2013).

The LC-ICLV model is a hybrid model that accounts for both the subjectivity of an individual's behaviour (attitudes, norms, preferences, or lifestyles) and the preference heterogeneity across the population by integrating the latent constructs within the LC structure (Hurtubia et al., 2014; Alizadeh et al., 2019). As one of the studies that have used the LC-ICLV model, Hurtubia et al. (2014) applied it to two mode choice case studies in Switzerland and reported that incorporating the psychometric indicators into the LC model can enhance the explanatory power of the LC model. The same conclusion was drawn by Alizadeh et al. (2019), who developed an LC-ICLV model to explain the differences in route choice behaviour between frequent and occasional drivers using observable variable and latent behaviour constructs.

### 2.3 Research Gaps and Contributions

The existing discrete choice literature revealed that hybrid choice modelling techniques have the potential to improve the explanatory power of traditional models. This improvement is achieved by incorporating individual's behavioural variables and/or accounting for the behavioural heterogeneity across the population's segments in the modelling process (Alizadeh et al., 2019; Hurtubia et al., 2014). Thus, results from hybrid choice models can provide useful inputs to policymakers in recognizing target market segments and proposing segment-specific recommendations pointed towards improving the attractiveness of the service (Saxena et al., 2020).

On the other side, the factors affecting the individual's willingness to use microtransit, ODT, and MOD transit vary from service to service. This variation is due to the differences in the operational characteristics among these services. For the ODT service, previous studies revealed that individual's willingness to use the service is affected by their sociodemographic characteristics as well as by the reliability, pickup and dropoff locations, and the cost of the service (Khattak and Yim, 2004; Yu et al., 2017). However, it is important to note that these studies rely on stated preference data, which reflect individuals' behaviour in a hypothetical scenario rather than their actual behaviour in the marketplace. The modelling techniques used in these studies could neither account for the subjectivity of human behaviour nor the preference heterogeneity among the market's latent segments in the decision-making process. Hence, their policy implications for transit operators are limited. Furthermore, to the best of the authors' knowledge, no previous studies have investigated the preferences of public transit users between FRT and ODT services, nor have advanced choice models (e.g., LC, ICLV, and LC-ICLV) been developed for ODT users to understand their preferences.

This study develops three different hybrid choice models, namely LC, ICLV, and LC-ICLV, to examine the preference of ODT users between FRT and ODT services. The hybrid choice models are developed in order to address the novel research questions outlined in the Introduction. In particular, the choice models examine the impact of both observable and latent variables on ODT captive and non-captive users' service preferences, based on which suggestions and practical implications are provided to assist transit operators in the design and operation of more convenient, attractive, and sustainable ODT services. Moreover, the hybrid choice models are estimated using a rich dataset that combines the actual level of service attributes and self-reported usage behaviour obtained for Belleville's ODT service.

## 3. Data Description

Two data sources were used to develop the fused data for this study: (a) a web-based revealed preference survey that was distributed to Belleville's ODT users through email in November 2019, and (b) operational data collected from Belleville's ODT pilot project from October till December 2018. It is worth mentioning that both datasets have been independently analyzed in previous studies. The revealed preference data were employed to describe the user profiles and explore the relationship between satisfaction with the ODT service and activity participation (Zhang et al., 2021 and Zhang et al., 2020). The operational data were used to perform spatiotemporal analysis of supply, level of service, and origin-destination patterns, as well as to develop trip production and distribution models (Sanaullah et al., 2021 and Alsaleh and Farooq, 2021). However, to better understand the ODT user preference between the FRT and ODT services, both datasets were merged and analyzed collectively in this study. In the revealed preference survey, the following information were collected for each participant:

- Sociodemographic and socioeconomic characteristics, for instance age, gender, marital status, education level, income, household size, and car ownership.
- Experience with the ODT service, such as trip purpose, in-vehicle time with the ODT service compared to the FRT service, and preferred mode at night between the FRT and ODT services.
- Travel history in the past few months, including the main travel mode before the ODT service was launched, and the main travel mode when the FRT was running at night.
- Attitudinal and perceptional questions about the ODT features. Participants were asked about their attitudes towards 7 statements, that were based on psychometric indicators, to capture their attitudes and perceptions towards the flexibility and quality of the ODT service, as shown in Table 1. More specifically, participants were asked to rate the importance of the waiting time, in-vehicle time, reliability, and the flexibility of ODT service on their decision to use the ODT service using a five-point Likert scale ranging from not at all important to extremely important. Besides, participants were asked to provide their satisfaction with the ODT mobile app interface, website interface, and their available services using a five-point Likert scale ranging from very unsatisfied to very satisfied.

**Table 1.** Attitudinal and perceptional questions

| No. | Indicator Type | Question |
|---|---|---|
| 1 | | How important is the waiting time in influencing your choice of using ODT at night? |

| **2** | Attitude towards the flexibility of ODT service | How important is the time on the ODT bus in influencing your choice of using ODT at night? |
|---|---|---|
| **3** | | How important is the reliability (bus shows up when it is supposed to) in influencing your choice of using ODT at night? |
| **4** | | How important are the convenience and the flexibility in influencing your choice of using ODT at night? |
| **5** | Perception towards the online services | How satisfied are you with the user interface of the ODT app? |
| **6** | | How satisfied are you with the user interface of the ODT website? |
| **7** | | How satisfied are you with the availability of schedule/maps/fares? |

The survey was sent via email to 1,342 users and 263 responses were received. For more details about the revealed preference survey used in this study, interested readers are referred to Zhang et al. (2020). On the other hand, the ODT service's operational data were collected from October to December 2018. These data contained detailed information about the trip's history for 430 users, including their IDs (email addresses), trip request creation date and time, trip request time, trip status (assigned or not assigned), actual pickup time, number of riders, pickup stop, and dropoff stop. For further details on the operational data analysis of Belleville's ODT service, we refer the readers to Sanaullah et al. (2021).

In this study, the revealed preference survey was used to obtain the socioeconomic and sociodemographic characteristics, psychometric indicators, self-reported attributes of the ODT service, and alternative travel modes of participants. Participant's choice preference between the ODT and the FRT services was based on their answers to the following statement: *"I would prefer to have fixed route busses at night"*. We considered those with *"Strongly Agree"* or *"Agree"* answers to prefer the FRT alternative, and those with *"Strongly Disagree"* or *"Disagree"* answers to prefer the ODT alternative. However, respondents who answered *"Neither agree nor disagree"* were considered to be indifferent between the FRT and ODT services. On the other hand, the operational data were used to obtain the number of assigned trips, the number of unassigned trips, and the average waiting time for the users. Then, the information obtained from both the revealed preference survey and the operational data of the ODT service were merged based on the common "email address" information, forming a fused dataset of 72 observations. It is worth noting that the differences between the fused dataset and both the operational and the revealed preference data were statistically verified. The results showed that there are no significant differences between the three samples at a 95% confidence level, indicating that the fused dataset is a representative sample of Belleville's ODT users (see Appendix A for details). Orme (1998) suggested the following equation to determine the minimum sample size for choice-based conjoint modelling:

$$N \geq (500 \times C/(T \times A)) \qquad \textbf{(1)}$$

where C is the largest number of any attributed levels, T is the number of choice tasks, and A is the number of alternatives. According to this rule of thumb, the minimum sample size required for our analysis is 52 respondents (N = (500 X 4) / (13 X 3) = 51.3).

In terms of the data protection aspect, user privacy is maintained by not revealing any disaggregate information publicly. Furthermore, participant's consent to use their data in the research was acquired during the survey by Zhang et al. (2020). The second dataset was extracted from the supply-demand matching and vehicle routing system used by the Belleville ODT service. In addition to the consent and authorization, user privacy was strictly maintained throughout the study. The use of email addresses was solely limited to matching the two datasets, while none of the information presented here can be traced back to the users. Table 2 provides a detailed description of the observed variables. The levels of assigned trips, unassigned trips, and the average waiting time variables, shown in Table 2, were based on the k-means clustering algorithm and elbow method results.

**Table 2.** Description of the observed variables

| Variable | N | % | Variable | N | % |
|---|---|---|---|---|---|
| **Socioeconomic and Sociodemographic** | | | Less than FRT | 19 | 26.4 |
| Age | | | Equal to FRT | 25 | 34.7 |
| Young (18 – 24) | 29 | 40.3 | More than FRT | 28 | 38.9 |
| Adults (25 – 44) | 33 | 45.8 | Trip Purposes | | |
| Middle-age (45 – 64) | 10 | 13.9 | Work-based | 29 | 40.3 |
| Old (more than 64) | 0 | 0 | Nonwork-based | 11 | 15.3 |
| Gender | | | Mixed purposes | 32 | 44.4 |
| Male | 34 | 47.2 | Travel mode when FRT was running at night | | |
| Female | 36 | 0.5 | Active Mode (walking or cycling) | 25 | 34.7 |
| Other | 2 | 2.8 | Car as a driver or passenger | 25 | 34.7 |
| Marital Status | | | FRT service | 14 | 19.5 |
| Single | 45 | 62.5 | Mobility Bus Service | 3 | 4.2 |
| Married | 9 | 12.5 | Not applicable | 5 | 6.9 |
| Widowed | 2 | 2.8 | **Actual Attributes** | | |
| Divorced | 5 | 6.9 | Assigned trips levels | | |
| Other | 11 | 15.3 | Low (less than 3) | 46 | 63.9 |
| Education Level | | | Medium (3 – 7) | 16 | 22.2 |
| Secondary School | 30 | 41.7 | High (more than 7) | 10 | 13.9 |
| Diploma | 24 | 33.3 | Unassigned trips levels | | |
| Undergraduate | 9 | 12.5 | Low (less than 5) | 46 | 63.9 |
| Graduate | 9 | 12.5 | Medium (5 – 11) | 16 | 22.2 |
| Annual Income Level (Thousand CAD) | | | High (12 – 23) | 8 | 11.1 |
| Low (less than 20) | 39 | 54.1 | Very High (more than 23) | 2 | 2.8 |
| Medium (20 – 50) | 22 | 30.6 | Average waiting time levels (minutes) | | |
| High (more than 50) | 11 | 15.3 | Low (less than 10) | 19 | 26.4 |
| Household Size Level | | | Medium (10 – 20) | 28 | 38.9 |
| Low (1 – 3) | 44 | 61.1 | High (21 - 33) | 20 | 27.8 |
| High (+4) | 28 | 38.9 | Very High (more than 33) | 5 | 6.9 |
| Car ownership | | | **Choice Alternatives** | | |
| 0 | 68 | 94.4 | FRT | 40 | 55.6 |
| +1 | 4 | 5.6 | ODT | 23 | 31.9 |
| **Self-reported Attributes** | | | | | |
| In-vehicle time with the ODT service | | | Indifferent | 9 | 12.5 |

## 4. Methodology and Model Specifications

To investigate the user preference, three hybrid choice models were developed, namely Latent Class (LC), Integrated Choice and Latent Variable (ICLV), and Latent Class Integrated Choice and Latent Variable (LC-ICLV) model. Moreover, a Multinomial Logit (MNL) model was used for comparison purposes. Hybrid choice models capture the impact of attitudes, perceptions, lifestyles, and market latent segments on user's preferences which can be of importance to derive policy measures that are more sustainable and have social as well as economic benefit compared with traditional discrete choice models. The variables used to formulate our models are described in Table 3. This section presents the framework as well as the four model specifications. However, readers can refer to (Alizadeh et al., 2019; Hurtubia et al., 2014) for the detailed econometric formulation of the LC, ICLV, and LC-ICLV models.

**Table 3.** Description of the transformed observed variables used in modelling

| Variable | Definition | Variable | Definition |
|---|---|---|---|
| *Choice* | =1 if the preference is FRT<br>2 for ODT<br>3 for indifferent | *NonworkTrip* | =1 if user used the ODT service for non-work purposes |
| *Male* | =1 if user's gender is male | *MixedTrip* | =1 if user used the ODT service for work and non-work purposes |
| *Young* | =1 if user is young | *FixedService* | =1 if user used FRT mode when it was running at night |
| *MiddleAge* | =1 if user is middle-aged | *ActiveMode* | =1 if user used walking or cycling modes when the FRT service was running at night |
| *LowIncome* | =1 if user's income level is low | *InVeh_less* | =1 if user reported that the in-vehicle time of the ODT trips is less compared to the FRT trips |
| *HighIncome* | =1 if user's income level is high | *InVeh_more* | =1 if user reported that the in-vehicle time of the ODT trips is higher compared to the FRT trips |
| *Hhld_L* | =1 if user is living in a household size of 3 or fewer people. | *Assigned_L* | =1 if user's assigned trips level is low |
| *Hhld_H* | =1 if user is living in a household size of 4 or more people. | *Assigned_H* | =1 if user's assigned trips level is high |
| *Single* | =1 if user's marital status is single | *Waiting_L* | =1 if user experienced a low waiting time level |
| *Sec_school* | =1 if user's highest education level is a secondary school | *Waiting_H* | =1 if user experienced a high waiting time level |
| *HigherEdu* | =1 if user has an undergraduate or graduate degree | *Unassigned_L* | =1 if user's unassigned trips level is low |
| *Car* | =1 if user has a car | *Unassigned_H* | =1 if user's unassigned trips level is high |
| *WorkTrip* | =1 if user used the ODT service for work purposes | | |

### 4.1 Multinomial Logit (MNL) Model

In the MNL model, the systematic component of the utility functions is defined using the observable variables only (Ababio-Donkor et al., 2020). Thus, the systematic utility functions of FRT, ODT, and indifferent alternatives were formulated using the socioeconomic and demographic characteristics of ODT users as well as the actual and the self-reported attributes of ODT service. The final specification of the MNL utility functions is presented in Appendix B.

### 4.2 Latent Class (LC) Model

Previous literature has shown that the choice preference behaviour varies and there may exist a significant heterogeneity across the population (Zhou et al., 2020; Alizadeh et al., 2019; Hurtubia et al., 2014). The LC model addresses this pattern by probabilistically dividing the population into certain latent classes through the definition of class membership functions and class-specific utility functions. Class membership functions are defined using individual characteristics to determine the probability of each individual belonging to a certain latent class. Class-specific utility functions are defined based on observable variables to account for individual preference heterogeneity across the population's latent classes (Alizadeh et al., 2019; Hurtubia et al., 2014).

In this study, the LC model was used to compare the service preferences between ODT captive and non-captive users. Public transit captive users are those who have no other option than using public transit service to meet their travel needs, as they do not have the financial capability to use other travel modes (Krizekand El-Geneidy, 2007). The clustering analysis based on previous trip history and income revealed the existence of these two classes in our case study (see Appendix C). Accordingly, the class membership functions were defined with a binary logit model structure based on the definition of public transit captive user (see Appendix D). For both latent classes, ODT captive and non-captive users, specific utility functions were developed using socioeconomic and demographic characteristics as well as ODT service self-reported and actual attributes (see Appendix D).

### 4.3 Integrated Choice and Latent Variable (ICLV) Model

It has been shown that attitudes, perceptions, and lifestyles influence the decision-making process (Zhou et al., 2020; Alizadeh et al., 2019; Hurtubia et al., 2014). To incorporate these unobservable (latent) variables into the systematic component of the utility functions, ICLV model is used. ICLV consists of a choice model and latent variable model components. The former component is used to capture the impact of observable variables on individuals' decision-making behaviour, while the latter incorporates latent variables in the estimation process. The definition process of the latter component involves three main steps: (1) identifying latent variables, (2) developing measurement equations for indicators, and (3) developing structural equations for the latent variables. To identify latent variables, the principal factor analysis method is applied to the psychometric indicators obtained from users' responses to behavioural questions. The results of this analysis reveal the unobservable variables that the indicators can represent and show the contribution of each indicator in explaining these variables. After that, measurement and structural equations are developed for the identified latent variables. In the measurement equations, each latent variable is linked to its observable indicators and, in the structural equations, latent variables are defined using observable variables (Ababio-Donkor et al., 2020; Alizadeh et al., 2019).

Once the choice and the latent variable components are defined, the systematic utility functions of the ICLV model can be developed by incorporating the latent variable model into the choice model. This integration can be performed in two different ways, depending on the purpose for which latent variables are used. Latent variables are added directly to the choice model when used to apprehend the influence of individuals' attitudes, perceptions, or lifestyles on their choice behaviour (see, for example, Ababio-Donkor et al., 2020). However, when a latent variable is used to capture its influence on a specific alternative's attribute, then it is incorporated in the choice model by multiplying that attribute either with latent variable (see, for example, Alizadeh et al., 2019) or with the exponential of the latent variable (see, for example, Bierlaire, 2018).

In this study, the ICLV framework was used to investigate the impact of user attitudes and perceptions towards the ODT features on their service preferences. The choice model component of the ICLV model was defined using sociodemographic characteristics as well as ODT service self-reported and actual attributes. On the other hand, the principal factor analysis method was applied to the psychometric indicators shown in Table 1, and the results revealed the presence of two distinct components (see Appendix E). The first component represents the ODT users whose choice of using the ODT service is subjected to the waiting time, in-vehicle time, reliability, and flexibility of the service. The second component corresponds to the ODT users who are satisfied with the ODT mobile app interface, the website interface, and their available services. We refer to the first component as the *Time Sensitivity* latent variable and the second component as the *Online Service Satisfaction* latent variable. The structural equations of the Time Sensitivity and Online Service Satisfaction latent variables were defined using socioeconomic and demographic characteristics. Appendix E contains the specification of the structural equations, the measurement equation for indicators, and the specification of the systematic utility functions for the three alternatives. It is worth mentioning that both latent variables were added to the systematic utility function of the ODT alternative to find out how the users' attitudes and perceptions towards the ODT features can affect their preference towards the ODT alternative.

4.4 Latent Class Integrated Choice and Latent Variable (LC-ICLV) Model

In the LC-ICLV model, the latent variable model is incorporated into the systematic utility functions of the LC model. Thus, the model accounts for both the subjectivity of human behaviour and the behavioural heterogeneity across the population's latent segments in the decision-making process. In this study, the LC-ICLV framework shown in Figure 2 was used to apprehend the influence of the observable variables as well as the latent variables on ODT captive and non-captive users' service preferences. The systematic utility functions of the LC-ICLV model were developed by integrating the latent variable model defined in Section 4.3 into the LC model defined in Section 4.2. The class membership functions are presented in Table 4 and the class-specific utility functions for alternatives are outlined in Table 5.

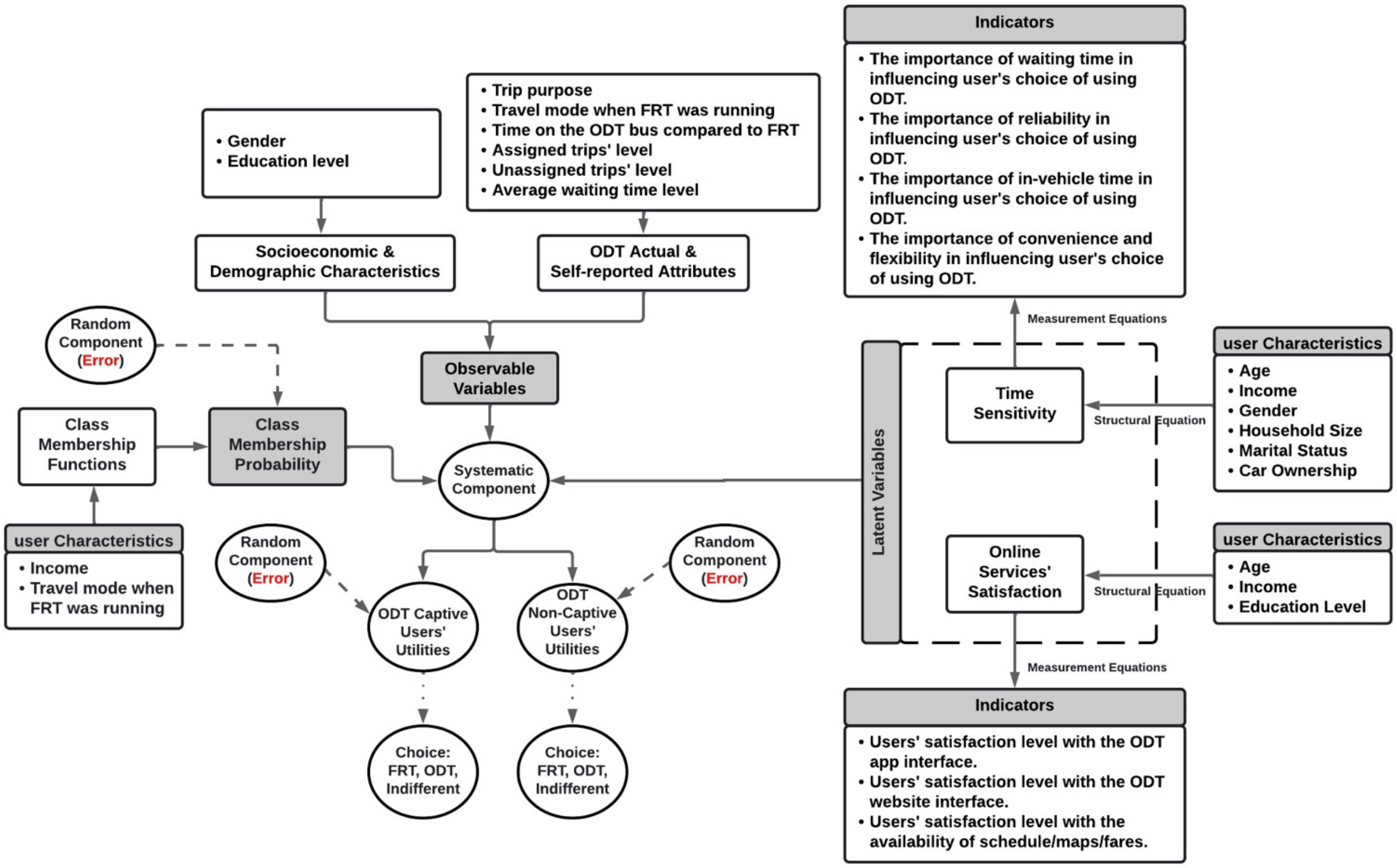

**Figure 2.** LC-ICLV modelling framework

**Table 4.** Class membership function specifications for the IC-ICLV model

| Parameter | Class 1 (Captive Users) | Class 2 (Non-Captive Users) |
|---|---|---|
| $\gamma_{CAP}$ | 1 | - |
| $\gamma_{INCOME}$ | $LowIncome$ | - |
| $\gamma_{MODE}$ | $FixedService$ | - |

**Table 5.** Latent Class Integrated Choice and Latent Variable (LC-ICLV) model specification

| Parameter | Class 1 (Captive Users) | | | Class 2 (Non-Captive Users) | | |
|---|---|---|---|---|---|---|
| | ODT | FRT | Indiff. | ODT | FRT | Indiff. |
| $ASC_{ODT}^{C1}$ | 1 | - | - | - | - | - |
| $ASC_{INDIFF}^{C1}$ | - | - | 1 | - | - | - |
| $\beta_{purpose}^{C1}$ | MixedTrip | NonworkTrip | - | - | - | - |
| $\beta_{in-veh}^{C1}$ | InVeh_less | InVeh_more | - | - | - | - |
| $\beta_{waiting}^{C1}$ | Waiting_L | Waiting_H | - | - | - | - |
| $\beta_{unassigned_trips}^{C1}$ | Unassigned_L | - | - | - | - | - |
| $\beta_{mode}^{C1}$ | ActiveMode | FixedService | - | - | - | - |
| $\beta_{edu}^{C1}$ | - | - | HigherEdu | - | - | - |
| $\beta_{gender}^{C1}$ | - | - | Male | - | - | - |
| $\beta_{TS}^{C1}$ | TIME_SEN | - | - | - | - | - |
| $\beta_{OSS}^{C1}$ | ON_SERV_SAT | - | - | - | - | - |
| $ASC_{ODT}^{C2}$ | - | - | - | 1 | - | - |
| $ASC_{INDIFF}^{C2}$ | - | - | - | - | - | 1 |
| $\beta_{purpose}^{C2}$ | - | - | - | NonworkTrip | WorkTrip | - |
| $\beta_{in-veh}^{C2}$ | - | - | - | InVeh_less | InVeh_more | - |
| $\beta_{waiting}^{C2}$ | - | - | - | Waiting_L | Waiting_H | - |
| $\beta_{assigned_trips}^{C2}$ | - | - | - | Assigned_L | Assigned_H | - |
| $\beta_{edu}^{C2}$ | - | - | - | - | - | HigherEdu |
| $\beta_{gender}^{C2}$ | - | - | - | - | - | Male |
| $\beta_{TS}^{C2}$ | - | - | - | TIME_SEN | - | - |
| $\beta_{OSS}^{C2}$ | - | - | - | ON_SERV_SAT | - | - |

## 5. Results

All models were estimated using PandasBiogeme software package in Python (Bierlaire & Fetiarison, 2009; Bierlaire, 2003), which is based on the maximum likelihood estimation technique. During the estimation process, we considered starting with simpler models and specifications to get good starting values and minimize the estimation cost of more complex models (Alizadeh et al., 2019; Hurtubia et al., 2014). The ICLV and the LC-ICLV models were estimated using the full information estimation method, described by Bierlaire (2018), to jointly estimate the parameters of the choice model, latent variable model, and the class membership functions (Alizadeh et al., 2019; Bierlaire, 2018). Tables 6 and 7 present detailed estimation results for the LC-ICLV model. The estimation results for the simpler specifications, i.e., MNL, LC, and ICLV models, are presented in Appendix F. In this section, we discuss the impacts of the observed as well as latent variables on the service preference.

**Table 6.** Estimation results of the measurement equations of the LC-ICLV model

| Indicator | Parameter | Estimate | Rob. t-test |
|---|---|---|---|
| WAIT_IMPO | $\alpha_{\text{WAIT_IMPO}}$ | 0.000 | – |
| | $\beta_{\text{WAIT_IMPO}}$ | 1.000 | – |
| | $\sigma_{\text{WAIT_IMPO}}$ | 1.000 | – |
| RELIA_IMPO | $\alpha_{\text{RELIA_IMPO}}$ | 1.200 | 2.54 |
| | $\beta_{\text{RELIA_IMPO}}$ | 0.785 | $1.73^{*}$ |
| | $\sigma_{\text{RELIA_IMPO}}$ | 1.520 | 3.31 |
| TIME_BUS | $\alpha_{\text{TIME_BUS}}$ | -0.058 | $-0.20^{**}$ |
| | $\beta_{\text{TIME_BUS}}$ | 0.607 | 2.17 |
| | $\sigma_{\text{TIME_BUS}}$ | 1.230 | 6.37 |
| FLEXIBILITY | $\alpha_{\text{FLEXIBILITY}}$ | 1.030 | 2.00 |
| | $\beta_{\text{FLEXIBILITY}}$ | 0.099 | $0.24^{**}$ |
| | $\sigma_{\text{FLEXIBILITY}}$ | 1.31 | 5.33 |
| APP_INTER | $\alpha_{\text{APP_INTER}}$ | 0.000 | – |
| | $\beta_{\text{APP_INTER}}$ | 1.000 | – |
| | $\sigma_{\text{APP_INTER}}$ | 1.000 | – |
| WEB_INTER | $\alpha_{\text{WEB_INTER}}$ | 0.121 | 1.42** |
| | $\beta_{\text{WEB_INTER}}$ | 0.833 | 7.92 |
| | $\sigma_{\text{WEB_INTER}}$ | 0.387 | 3.17 |
| AVAIL_SERV | $\alpha_{\text{AVAIL_SERV}}$ | 0.388 | 3.1 |
| | $\beta_{\text{AVAIL_SERV}}$ | 0.584 | 3.88 |
| | $\sigma_{\text{AVAIL_SERV}}$ | 0.766 | 7.05 |

* Not statistically significant at 95% confidence level
** Not statistically significant at 90% confidence level

**Table 7.** Estimation results of the choice model, latent variable model, and class membership functions of the LC-ICLV model

| **Parameters** | **Estimate** | **Rob. t-test** | **Parameters** | **Estimate** | **Rob. t-test** |
|---|---|---|---|---|---|
| **Choice Model** | | | $\alpha_{CONS}^{TS}$ | 0.266 | 0.69** |
| $ASC_{ODT}^{C1}$ | -11.600 | -5.30 | $\alpha_{age}^{TS}$ | 1.150 | 2.40 |
| $ASC_{INDIFF}^{C1}$ | -2.540 | -2.18 | $\alpha_{income}^{TS}$ | 0.617 | 1.80* |
| $\beta_{purpose}^{C1}$ | 0.860 | 1.53** | $\alpha_{car}^{TS}$ | -0.485 | -0.93** |
| $\beta_{in-veh}^{C1}$ | 1.600 | 2.20 | $\alpha_{hhld}^{TS}$ | 0.684 | 1.90* |
| $\beta_{waiting}^{C1}$ | 0.377 | 0.66** | $\alpha_{gender}^{TS}$ | 1.130 | 2.85 |
| $\beta_{unassigned_trips}^{C1}$ | 10.400 | 7.50 | $\alpha_{marital}^{TS}$ | -0.979 | -2.01 |
| $\beta_{mode}^{C1}$ | 1.37 | 2.01 | $\sigma^{TS}$ | -0.224 | -0.91** |
| $\beta_{edu}^{C1}$ | 2.31 | 1.82* | $\alpha_{CONS}^{OSS}$ | -0.470 | -1.58** |
| $\beta_{gender}^{C1}$ | 0.943 | 0.67** | $\alpha_{age}^{OSS}$ | 0.991 | 2.31 |
| $\beta_{TS}^{C1}$ | -0.138 | -0.17** | $\alpha_{income}^{OSS}$ | 0.600 | 2.12 |
| $\beta_{OSS}^{C1}$ | 0.053 | 0.10** | $\alpha_{edu}^{OSS}$ | 0.605 | 1.69* |
| $ASC_{ODT}^{C2}$ | - | - | $\sigma^{OSS}$ | 1.14 | 5.20 |
| $ASC_{INDIFF}^{C2}$ | - | - | **Class Membership Functions** | | |
| $\beta_{purpose}^{C2}$ | 50.000 | 3.88 | $\gamma_{CAP}$ | -10.700 | -40.00 |
| $\beta_{in-veh}^{C2}$ | 27.700 | 3.66 | $\gamma_{INCOME}$ | 24.100 | 5.68 |
| $\beta_{waiting}^{C2}$ | 6.010 | 1.34** | $\gamma_{MODE}$ | 21.600 | 34.10 |
| $\beta_{assigned_trips}^{C2}$ | 10.900 | 3.47 | **Performance Indicators** | | |
| $\beta_{edu}^{C2}$ | 10.600 | 3.30 | Number of observations | 72 | |
| $\beta_{gender}^{C2}$ | 36.100 | 4.07 | Number of parameters | 50 | |
| $\beta_{TS}^{C2}$ | 9.440 | 2.27 | Initial log-likelihood | -951.43 | |
| $\beta_{OSS}^{C2}$ | 5.630 | 2.17 | Final log-likelihood | -567.44 | |

| **Latent Variable Model (Structural Equations)** | Rho-square-bar | 0.351 |
|---|---|---|

[*] Not statistically significant at 95% confidence level

[**] Not statistically significant at 90% confidence level

## 5.1 Impact of Observed Variables on User Service Preference

Here we discuss the impact of user's characteristics and attributes of alternatives on the service preference decision. As all the models are giving us consistent parameter values, we are able to generalize various trends.

### 5.1.1 General User Preferences

Figure 3 summarizes the impact of observed variables on service preference. All else equal, the FRT alternative is generally the preferred alternative of ODT users, followed by the indifferent alternative. It is worth mentioning that the performance of Belleville's ODT service was unstable during the first three months of its operation, from October till December 2018. As a result, waiting time was higher compared with the first three months of 2019 (Sanaullah et al., 2021). Our results suggest that the performance of the service in this period had a significant influence on preferences. Users have a higher tendency to prefer the ODT alternative, if they took fewer trips and experienced shorter waiting time. However, users are more likely to prefer the FRT alternative if they took more trips and experienced higher waiting times. Hence, operators delivering new ODT projects should give more attention to the early-stage performance of the service to make it more convenient and attractive.

In addition, the trip purpose, main mode before the ODT service started, and in-vehicle travel time using the ODT service had significant impact on user preference. Users are more likely to prefer the ODT alternative if they use the ODT service mainly for non-work purpose, used non-motorized travel modes before the ODT service started, and their in-vehicle travel time with the ODT service is less when compared to the FRT service. Whereas users who used the ODT service mainly for commuting to and from work, used the FRT service before the ODT service started, and whose in-vehicle travel time after converting the service became longer, they had a higher tendency to prefer the FRT alternative. It is noted that being a middle-aged user and living in a large household increase the preference for ODT alternative. Moreover, males and highly educated users, who had an undergraduate or graduate degree, are most likely to be indifferent between ODT and FRT services. This could be either due to the availability of alternative travel modes for this category or their limited overnight trips that, in return, influenced their preferences.

### 5.1.2 Captive and Non-captive User Preferences

Figure 4 describes the impact of observed variables on the preferences of captive and non-captive users. It is observed that the probability of a user belonging to Class 1 increases with having low-income and with FRT service being the main travel mode before the ODT service started. As pointed out earlier, the clustering analysis based on previous trip history and income revealed the existence of these two classes in our data. Users belonging to Class 1 are entitled as "Captive Users", as their characteristics are consistent with the definition of public transit captive users described in Krizek and El-Geneidy (2007). Accordingly, individuals in Class 2 are considered as "Non-captive Users".

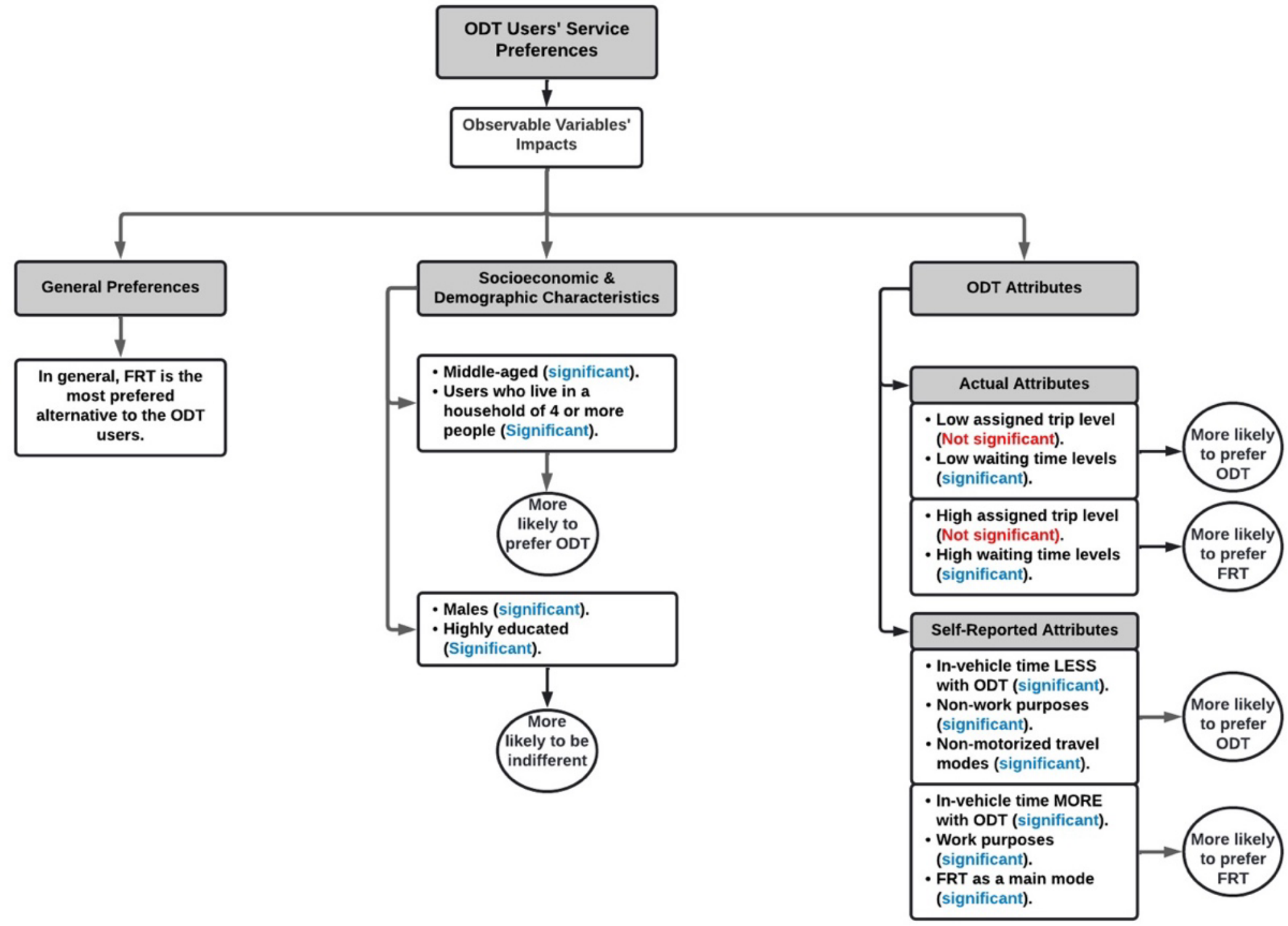

**Figure 3.** Impact of observed variables on service preference

Captive user's service preferences are most affected by the number of their unassigned trips during the early stages of the service. Users are more likely to prefer the ODT alternative if they encountered a low level of unassigned trips. Additionally, in-vehicle travel time had the second-highest impact on captive user's service preference. Captive users had a higher tendency to prefer the ODT alternative if the service had reduced their in-vehicle times in comparison to FRT. However, they are more likely to prefer the FRT alternative if the new service increased their in-vehicle times. In accordance with these results, captive users generally prefer the FRT service more than the ODT alternative, most likely due to their negative experience with the ODT service, especially during the early stages of the service. It is also seen that user's primary mode before the ODT started had a significant impact on the preference. Users who used non-motorized travel modes were more likely to prefer the ODT alternative, while those who used the FRT service had a higher tendency to prefer the FRT service. This indicates that ODT trips are likely to substitute for non-motorized trips the captive users make. Furthermore, highly educated captive users are indifferent between ODT and the FRT services, possibly due to their limited over-night trips (average trips for the category was 1.3, compared to 2.79 for the sample) that, in return, influenced their preference.

On the other hand, there is a significant difference in the service preference between captive and non-captive users. The service preferences of non-captive users are most affected by the trip purpose. Non-captive users are more likely to prefer the ODT alternative if they use the service to perform non-work trips. This indicates that the ODT service has the potential to satisfy non-captive user's needs to engage in various activities across the city, due to its flexibility and the large coverage area. However, non-captive users have a higher tendency to prefer the FRT alternative if their use is limited to work—probably because their travel time increased using the ODT service.

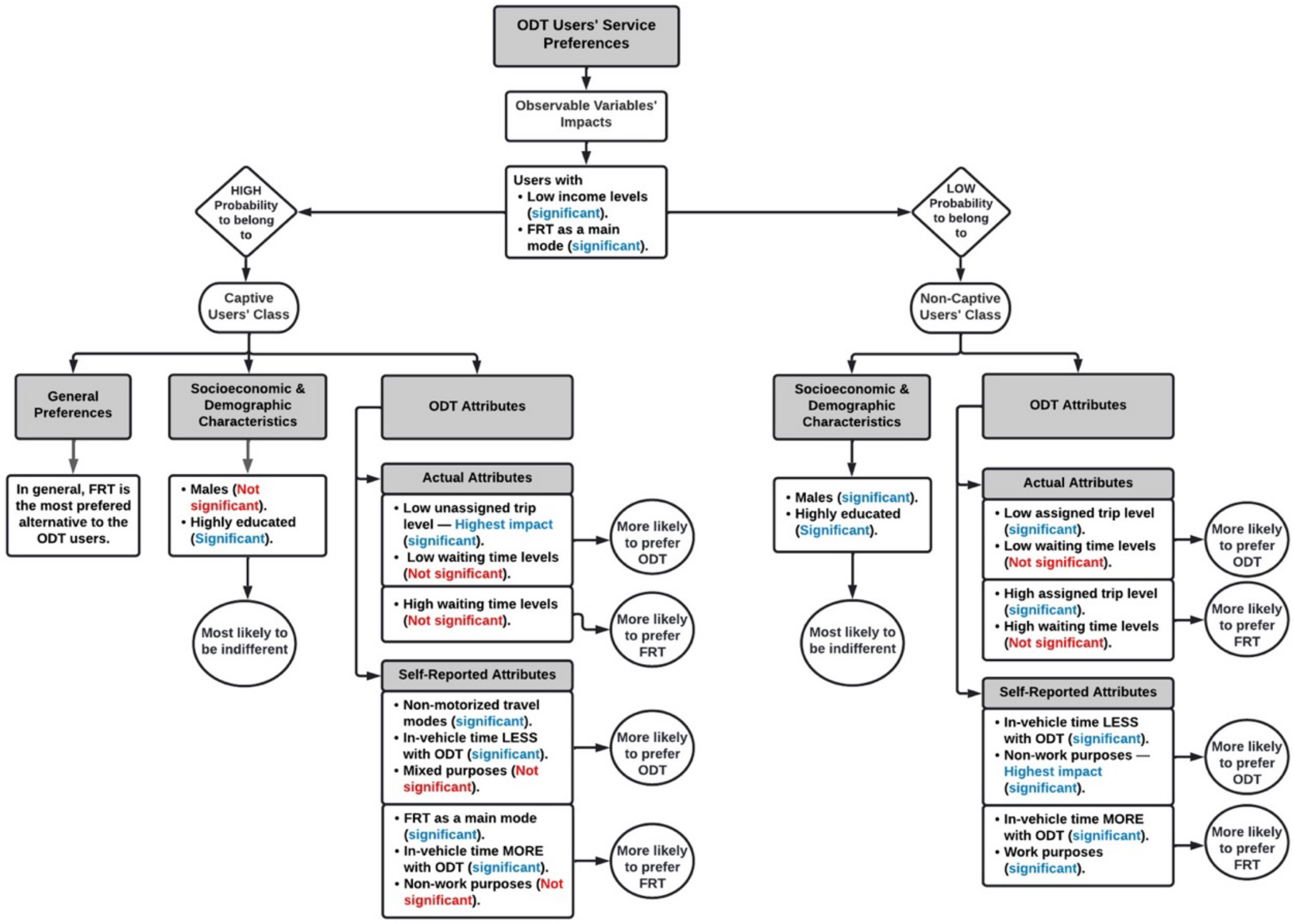

**Figure 4.** Impacts of observed variables on the preferences of captive and non-captive ODT users

Moreover, non-captive users who took a high number of trips in the first three months of the operation and experienced a high waiting time are more likely to prefer the FRT alternative. In contrast, non-captive users have a higher tendency to prefer the ODT alternative if they took fewer trips and experienced a low waiting time. These findings further stress the influence of the early-stage performance of the ODT service on convenience and attractiveness. It can also be noted that the in-vehicle time has a significant impact on non-captive user's service preferences. Non-captive users are more likely to prefer the ODT alternative if their in-vehicle time is less compared with the FRT service. Furthermore, male and highly educated users are indifferent between ODT and the FRT services. This may be because of the availability of alternative travel modes.

It can be noted that the results support our hypothesis about the latent classes. The preference of captive users is most affected by the unassigned trips level, whereas the purpose of the trip and the performance of the service have the highest impact on the non-captive users' preference.

## 5.2 Impact of Latent Variables on User Service Preference

The Time Sensitivity latent variable is positively correlated with the importance of the waiting time, in-vehicle time, and the reliability of the service (see Table 6). The Online Service Satisfaction latent variable is positively correlated to the user's satisfaction with the ODT mobile app interface, website interface, and the available features.

Figure 5 outlines the estimation results for the structural equations of the latent variables and shows their impacts on ODT user's service preferences. It is seen that the Time Sensitivity attitude is associated with being a male, young user, having a high income, and living in a household size of 3 or less. This is a logical finding, as it represents the category of people who usually participate in different recreational activities at night and can use alternative travel modes. Hence, their decision to use the ODT service is subjected to waiting time, in-vehicle time, and the reliability of the service. In contrast, single users are less likely to have the Time Sensitivity attitude, possibly because they have fewer responsibilities and more free time than other users. It is also observed that the Online Service Satisfaction latent variable corresponds to middle-aged users who have low income and a secondary school degree level education. We believe that this category may represent users who have not used any app-based ride-hailing services before, given that they generally tend to be unsatisfied with the online services provided to them. These findings suggest that the online services of the current ODT system need further improvements.

### 5.2.1 General User Preferences

As Figure 5 illustrates, time-sensitive users are more likely to prefer the ODT alternative. Similarly, users who are satisfied with the online service provided by the ODT mobile app and website have a higher tendency to prefer the ODT alternative. However, the parameter estimates of both latent variables, are insignificant at 95% confidence level (t = 1.29 and 1.12, respectively). The insignificance of these parameters is most probably due to the smaller sample size, which is a limitation of this study. Nevertheless, the inclusion of the latent variables in the MNL model improved its model fit (Rho-square-bar = 0.299 for ICLV model compared to 0.26 for the MNL model).

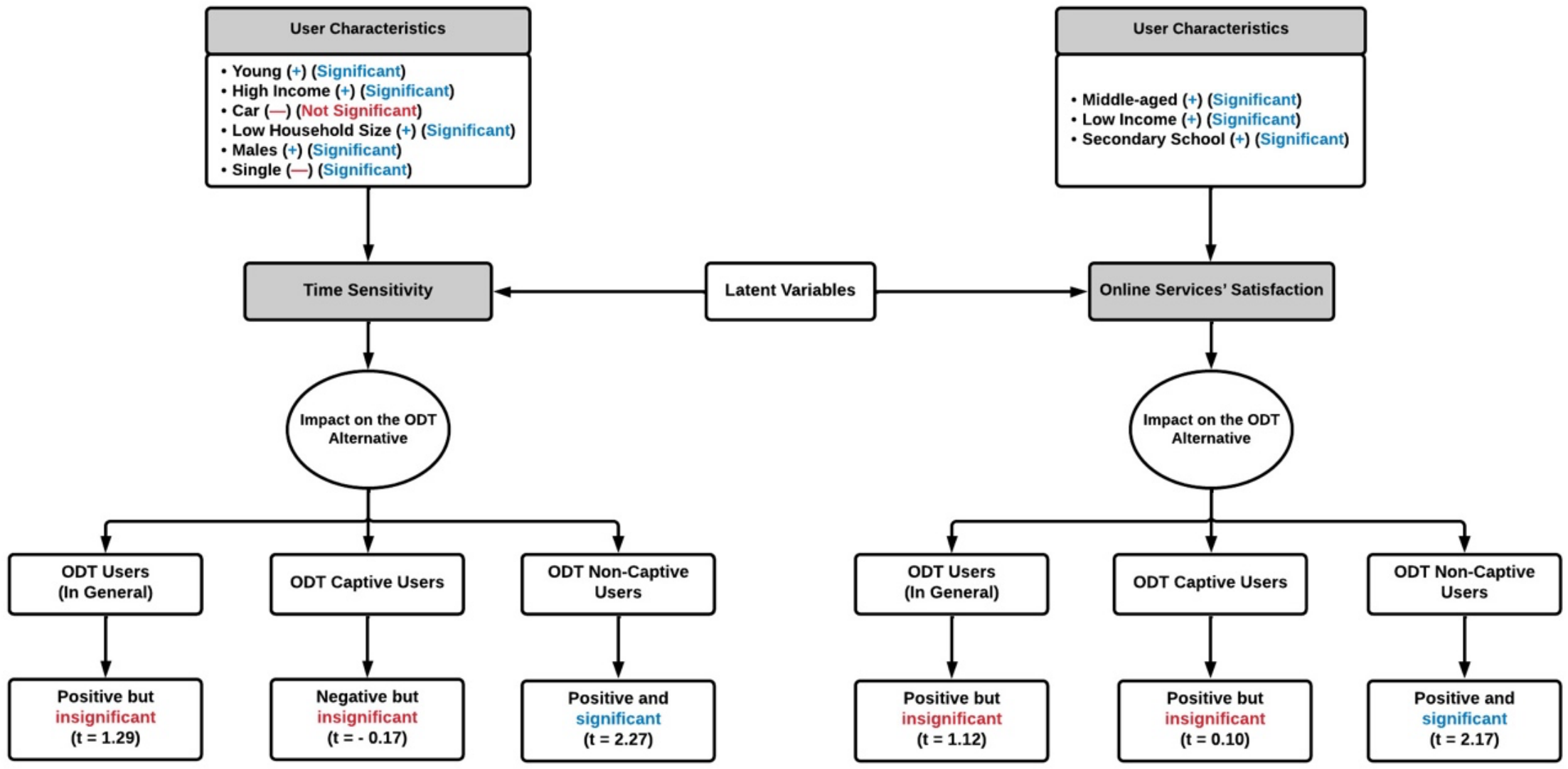

**Figure 5.** Impact of Latent variables on the ODT user's service preferences

### 5.2.2 Captive and Non-captive User Preferences

The Time Sensitivity and Online Service Satisfaction latent variables have insignificant impacts on the service preferences of ODT captive users. This is expected since the captive users class represents the type of users who have no other choice than accepting the new service. On the flip side, both latent variables have a significant positive impact on non-captive user's preference for the ODT alternative. Non-captive users who have the Time Sensitivity attitude are more likely to prefer the ODT alternative, indicating that their experience with the ODT service was positive. Similarly, non-captive users have a higher tendency to prefer the ODT alternative if they are satisfied with the online services provided by the ODT mobile app and website. These findings further support our hypothesis about the latent variables. Furthermore, the results suggest that improving the functionality of the service app and website might make ODT service more attractive.

## 5.3 Scenario Analysis

This section examines the potential impacts of improving the quality of the ODT service on user preferences based on estimation results from the MNL and LC-ICLV models. Three different scenarios are analyzed, each capturing the effect of modifying a specific attribute of the ODT service on user preferences. In the first scenario, we examine the impact of changing the distribution of waiting time categories to 50%, 25%, 20%, and 5%, respectively, for low, medium, high, and very high waiting times (originally, the distribution was 26.4%, 38.9%, 27.8%, and 6.9%). The second scenario explores the impact of modifying the distribution of in-vehicle travel time categories to 50%, 30%, and 20%, respectively, for less than FRT travel time, equal to FRT travel time, and greater than FRT travel time (originally, these categories were 26.4%, 34.7%, and 38.9%). In the final scenario, we consider the combined effect of improving both waiting and in-vehicle travel times by simultaneously implementing the modifications from the first and second scenarios. It is worth noting that these improvements could be achieved by operating medium-sized buses instead of the existing 40-ft buses. Additionally, the MNL and LC-ICLV models are applied to the original dataset to represent the do-nothing scenario, where no improvements are made to the ODT service. To gain a comprehensive understanding of the impacts of these scenarios, we applied the models to a larger synthesized dataset with 1000 observations. The dataset was generated using the Monte Carlo simulation technique, which ensures that the statistical properties of the original data are preserved.

Figure 6 illustrates the outcomes of the scenarios under consideration. The MNL results indicate that, in the do-nothing scenario, 270 users (27%) prefer the ODT service, 650 users prefer the FRT service, and 80 users are indifferent. Improving the waiting time for ODT leads to a 17.8% increase in users' preference for the service, which translates into 48 additional users favouring ODT over FRT. The shift in preference is a result of a 5.8% decrease in the number of users (38 users) preferring the FRT service and a 12.5% decrease in indifferent users (10 users). Enhancing the in-vehicle travel time of ODT yields a 28.5% increase in users' preference for the service, attracting 77 more users to prefer ODT over FRT. Additionally, there is a 10% increase in indifferent users (8 users) and a 13.1% decrease in users (85 users) favouring the FRT service. Combining improvements in both waiting and in-vehicle travel times results in a 47% increase in users' preference for the ODT service, with 127 more users favouring ODT over FRT. This comprehensive enhancement also leads to a notable 19.1% decrease in users (124 users) preferring the FRT service and a slight 3.8% decrease in indifferent users (3 users).

The do-nothing scenario of the LC-ICLV model yields similar results to those of the MNL model with only slight differences. The ODT service is preferred by 299 users (29.9%), while the FRT service is preferred by 643 users, and 58 users are indifferent to the service. However, improving the waiting time for ODT only results in a 1.7% increase in users' preference for the service, translating into 5 additional users preferring ODT over FRT. This is primarily due to the fact that most users (63%) belong to Class 1 (Captive Users), where waiting times have an insignificant effect on user preferences. On the other hand, improving ODT's in-vehicle travel time leads to an increase of 31.7% in users' preference for the service, attracting 95 more users to choose ODT over FRT. Furthermore, it results in a 12.9% decrease in FRT users (83 users) and a 24.1% decrease in indifferent users (14 users). It is mainly due to the fact that the in-vehicle travel time has the second greatest impact on user preferences in Class 1. Finally, combining improvements in waiting and in-vehicle travel times yields similar results to the second scenario. The combined improvements result in a 36.5% increase in users preferring the ODT service, a 15.2% decrease in users preferring the FRT service, and a 19% decrease in indifferent users.

It is important to note that the differences in scenario results obtained from the MNL and LC-ICLV models are primarily due to their inherent characteristics. The MNL model offers insights into general user preferences, while the LC-ICLV model provides market-specific preferences. This distinction highlights the importance of utilizing hybrid choice models to gain a comprehensive understanding of market needs. Furthermore, it emphasizes the benefits of using these advanced models for the development of targeted policy recommendations, effective marketing strategies, and designing service improvements tailored to the unique preferences of each market segment.

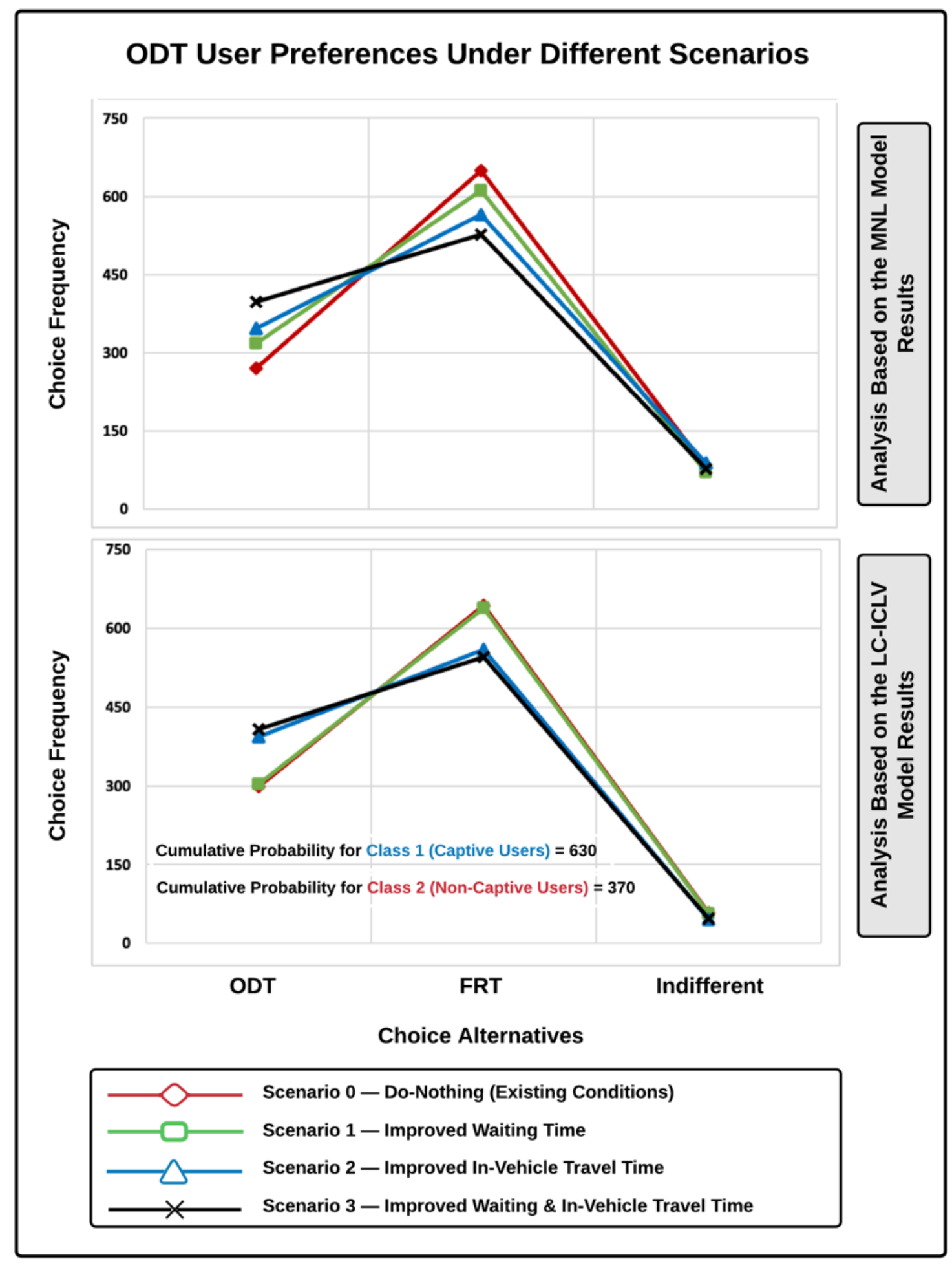

**Figure 6.** Impact of ODT service improvement Scenarios on user preferences

## 6. Discussion and Recommendations

This study modelled the impact of observed as well as latent variables on the preference between ODT and FRT services. The results are based on a rich dataset that contains actual level of service attributes and self-reported usage behaviour for ODT users in Belleville. The combination of the operational and revealed preference datasets provided us with the opportunity to examine the impact of users' actual experience with Belleville's ODT service (i.e., waiting time, assigned trips, and unassigned ride requests) on their preferences. The results would not be as informative if only revealed preferences were used, since participants' self-reported experiences may lack accuracy and may be biased in favour of their prior preferences. Moreover, the modelling dataset only included users who had used the ODT service at least once. Thus, the service preference analysis performed in this study was based on the actual experience of users with both FRT and ODT services. It should be noted, however, that the dataset did not include users who used the FRT but did not use the ODT, which prevented us from assessing their service preferences as well as their main concerns and reasons for not using the service. The absence of such analysis is one of the limitations of this study and may be an interesting area for future research.

The study has another limitation in that the sample size used for modelling is small. ODT services are still in their infancy with only a limited number of projects already in place. Additionally, ODT projects are primarily provided in low-density areas, which further limits the number of users. Therefore, obtaining revealed or stated preference data from ODT users can be a particularly challenging endeavor, and comparing the sample size obtained for ODT services with that obtained for other modes of transportation operating in various settings and locations (e.g., ridesourcing services) would be unfair. However, this does not negate the need to assess the suitability of utilizing the resulting data for modelling purposes. In this regard, the reduced sample in this study underwent several checks before being used to develop choice models for ODT users. From the results, we can conclude that there were no statistical differences between the reduced and the full samples, and that the reduced sample met the minimum sample size required for choice-based conjoint modelling according to the rule of thumb suggested by Orme (1998). For estimation of hybrid choice models, we considered starting with simpler models in order to obtain appropriate initial values, a suitable specification, and reduce estimation time for more complex models. Besides, the full estimation approach, described by Bierlaire (2018), was used for the development of the ICLV and LC-ICLV models to jointly estimate the parameters of the choice model, latent variable model, and the class membership functions. In light of all of the above, the estimation results demonstrated consistency across the developed choice models, in terms of parameter values, signs, and significance, indicating that the data used in the modelling process were capable of producing statistically reliable results.

In this section, we compare our findings with the existing studies and outline the main recommendations that can be derived from our results.

## 6.1 Discussion

One of the main factors that has been used to investigate the user's preference for flexible transit is the performance of service. The study conducted by Khattak and Yim (2004) found that traveller's willingness to use a hypothetical ODT service is most affected by the reliability of the service. A recent study in Ontario, Canada, found that ODT user's willingness to engage in activities is most affected by their satisfaction with the reliability and the performance of the service (Zhang et al., 2021). In line with these studies, we found that the Time Sensitivity latent variable, which represents user's attitude towards the reliability, in-vehicle time, and waiting time of the ODT service, has a significantly positive impact on non-captive user's preference for the ODT alternative. We also found that the service preference of captive users is most affected by the performance of ODT service, especially during the early stages in particular.

User's socioeconomic and demographic characteristics are also found to be important factors in explaining their preferences. A recent study in Michigan, USA, showed that males and highly educated travellers are more likely to prefer MOD transit service (Yan et al., 2019b). Another recent study in Sydney, Australia, found that individuals with work-based trips have a higher uptake rate for MOD transit service than those with non-work trips (Saxena et al., 2020). On the contrary, we found that males and highly educated users are indifferent between ODT and FRT services. Furthermore, the users taking work-based trips had a higher tendency to prefer FRT service, while for non-work trips they were more likely to choose ODT service. We believe that this is probably due to the differences in the operational characteristics between the two services, as discussed in Section 2.1. Moreover, Jittrapirom et al. (2019) found that unfamiliarity with the microtransit service app is one of the main barriers for the elderly to use the service. Leistner and Steiner (2017) also reported that unfamiliarity with smartphones and apps is the main barrier for the elderly to use app-based transportation services. However, we found that the Online Service Satisfaction latent variable, which measures user's satisfaction with the ODT mobile app interface, website interface, and available features, has a significantly positive impact on non-captive user's preference. Thus, user's satisfaction with online services is another important factor that should be considered when studying their preferences for any app-based mobility services. It is important to note that the impact of Time Sensitivity and Online Service Satisfaction latent variables, as well as the early-stage performance of the ODT service (including assigned trips, unassigned trips, and waiting time), have not been investigated in the literature before.

Although Belleville's ODT is a late-night service, it is primarily used for commuting to and from work, thus allowing the findings to be generalized to other ODT services. The findings of this study are mainly applicable to ODT services operating in low-density areas of North America, since they have similar structures, user characteristics, demand patterns, transportation infrastructure, land-use patterns, as well as some demographic characteristics (e.g., median income and average household size). Nevertheless, the results may be applicable to other areas of the world that share similar characteristics with the City of Belleville in terms of user and area-wide characteristics.

The use of hybrid choice models was found to have several advantages in understanding the service preferences of ODT users between the ODT and FRT services. Employing the ICLV model, we investigated the impact of Time Sensitivity and Online Service Satisfaction on ODT users' service preferences. Including these latent variables in the MNL model significantly improved its explanatory power (see Table 8), consistent with findings from previous studies (Ababio-Donkor et al., 2020; Kamargianni et al., 2015). Additionally, the Latent Class (LC) model effectively explained the heterogeneity of service preferences among ODT captive and non-captive users, demonstrating a higher goodness of fit compared to the MNL model, consistent with existing literature (Greene and Hensher, 2013; Hess et al., 2009; Shen, 2009; Massiani et al., 2007). Incorporating the latent variables into the LC model further enhanced its explanatory power, allowing for the investigation of the impact of Time Sensitivity and Online Service Satisfaction on ODT captive and non-captive users.

Notably, the ICLV model exhibited a higher goodness of fit than the MNL model, and the LC-ICLV model outperformed the LC model. However, as Vij and Walker (2016) emphasized, the use of latent variables should not lead to better goodness of fit than simpler models without latent variables. Latent variables can eventually be replaced by observable variables through the definition of structural equations. Considering the choice component only, it was observed that the ICLV and LC-ICLV models exhibited the same goodness of fit as their reduced forms (MNL and LC, respectively). The higher goodness of fit observed in ICLV and LC-ICLV models when considering both the choice and latent variable components can be attributed to the structural equations for latent variables incorporating observable variables (e.g., income, marital status, and education) that are not included in simpler models (MNL and LC). Furthermore, it is important to note that the purpose of using hybrid choice models in this study is not to achieve a higher degree of goodness of fit than traditional models. Rather, this study seeks to understand the influence of observable and latent variables on the service preferences of different market segments in order to assist transit agencies in designing and operating more convenient, attractive, and sustainable ODT services.

On the other hand, developing hybrid choice models encounters some challenges. During the estimation process of the LC models, several trials had to be performed to find the final structure for the class membership function. In addition, both the ICLV and the LC-ICLV models need to be estimated using the full information estimation method, described by Bierlaire (2018), to jointly estimate the parameters of the choice model, latent variable model, and the class membership functions. In this approach, it is recommended to start with simpler models and specifications in order to get good starting values and minimize the estimation cost of the complex models. However, it is computationally expensive and difficult to obtain, especially when two or more latent variables are used (Bierlaire, 2018). Furthermore, multiple trials were performed in order to find the final form of the structural equations for the latent variable model component of the ICLV and LC-ICLV models.

**Table 8.** Performance comparison between MNL, LC, ICLV, and LC-ICLV models

| Components | Performance Measures | MNL | LC | ICLV | LC-ICLV |
|---|---|---|---|---|---|
| Choice and Latent Variable | Rho-square-bar | 0.26 | 0.286 | 0.299 | 0.351 |
| | Initial log-likelihood | -79.1 | -79.1 | -882.69 | -951.43 |
| | Final log-likelihood | -47.49 | -36.47 | -577.93 | -567.44 |
| | No. of parameters | 11 | 20 | 41 | 50 |
| Choice Only | Rho-square-bar | 0.26 | 0.286 | 0.26 | 0.287 |
| | Initial log-likelihood | -79.1 | -79.1 | -79.1 | -79.1 |
| | Final log-likelihood | -47.49 | -36.47 | -47.49 | -38.36 |
| | No. of parameters | 11 | 20 | 11 | 18 |
| | Features | - | Compare the service preferences of ODT captive and non-captive users | Investigate the impact of the Time Sensitivity and the Online Services' Satisfaction on ODT users' service preferences | Investigate the impact of the Time Sensitivity and the Online Services' Satisfaction on ODT captive and non-captive users' service preferences |

6.2 Design and Operational Recommendations

There are several useful recommendations that can be drawn from this study, which can be useful for agencies who intend to deliver new ODT projects as well as to enhance the performance of the ongoing ODT projects. Our findings show that the Time Sensitivity and Online Service Satisfaction latent variables have a significant positive impact on non-captive user's preference for the ODT service. Therefore, transit operators may consider these aspects when planning, designing, and operating new as well as ongoing ODT projects to significantly affect the choice-making process of new users and consequently increase the ridership of the service.

Our results related to the service preference of the captive and non-captive users provide key insights to transit operators to design and operate more convenient and attractive ODT services. We found that the service preferences of captive users are most affected by their number of unassigned trips and in-vehicle time with the ODT service. Hence, transit operators should continuously update the operating fleet size based on the real-time spatio-temporal demand the system receives in order to minimize the number of unassigned trips. Furthermore, user's in-vehicle and waiting times could be reduced by substituting the current 40-ft buses with smaller size vehicles, for instance minibuses or vans. This will help to minimize the number of detours made by a vehicle. It is worth mentioning that Sanaullah et al. (2021) also recommended that operating smaller size vehicles for ODT systems could enhance their performance. However, this could result in an increased labour cost associated with the operations, therefore, further study is needed for better understanding of the financial implication.

On the other hand, we found that the service preferences of non-captive users are mainly affected by their satisfaction with the online services, the performance of the ODT service, and their trip purposes. Non-captive users who were satisfied with the online services had a positive experience with the ODT service in terms of in-vehicle and waiting times, and use the service for non-work trips, had a higher tendency to prefer the ODT service. In contrast, those who were unsatisfied with the online services had a negative experience with the ODT service or used the service for work-related trips, were more likely to prefer the FRT service. In light of these findings, we provide the following suggestions to transit operators to further enhance the attractiveness of ODT services:

- Transit operators should consider improving the user-interface and the functionality of the service app and website.
- Agencies wanting to deliver new ODT projects should give more attention to early-stage performance, as it has a significant impact on non-captive user's service preferences.
- Transit operators should consider medium occupancy vehicles instead of using the 40-ft buses in order to minimize the in-vehicle and the waiting times of ODT users. However, in cases where small municipalities already have a fleet of high-capacity buses and they cannot afford to invest in smaller vehicles, the ride-matching algorithm should give priority to minimization of wait times and detours, rather than maximization of capacity utilization.
- To make ODT services more efficient and convenient for work trips, we suggest serving such trips independently from non-work trips. In other words, transit operators should ask users to provide their trip purpose while requesting their trips and dedicate a predefined number of vehicles for work trips. However, a simulation-based study is needed to verify the feasibility and the validity of this suggestion, as well as to determine the required vehicle type, fleet size, service area, and cost to operate such a service.

## 7. Conclusions

This study investigated the impact of sociodemographic, performance, and latent variables on the preference between the ODT and FRT service of captive and non-captive users. The service preference heterogeneity is untangled to help transit operators design, plan, and operate more convenient and attractive ODT service. Previous studies revealed that the hybrid choice modelling techniques can improve the explanatory power of the traditional discrete choice models by incorporating the individual's behavioural variables and/or accounting for the behavioural heterogeneity across the population's latent classes in the modelling process (Alizadeh et al., 2019; Hurtubia et al., 2014). Therefore, the study modelled the ODT user preferences using three hybrid choice models, namely ICLV, LC, and LC-ICLV models. The data used in the modelling process is a rich dataset that contained actual level of service attributes and self-reported usage behaviour for 72 users of Belleville's ODT service. Although the sample size is small, it is a representative sample for Belleville's ODT users as there are no significant differences between the reduced and the full samples. Moreover, according to the rule of thumb proposed by Orme (1998), the minimum sample size required for our analysis is 52 respondents.

The study showed that the preference for ODT service of captive users was significantly affected by the number of unassigned trips, in-vehicle time, and primary travel mode before the ODT started. However, the number of unassigned trips had the highest impact on their preferences. The results also indicated that the ODT service is likely to replace non-motorized modes. On the other hand, the preference of non-captive users was significantly affected by the ODT service performance and trip purpose. In terms of latent variables, the Time Sensitivity and Online Service Satisfaction variables had a significant positive impact on non-captive user's preference for the ODT service. By considering these variables during the planning, designing, and operations of new as well as ongoing ODT projects, one can expect a significant positive effect on the users and their choice-making process, resulting in the increased ridership of the service. An important finding that has emerged from this study is that non-captive user's preference for the ODT services is most affected by their trip purpose.

We provide important recommendations to the agencies wanting to deliver new ODT projects as well as to enhance the performance of the ongoing projects. The results suggest that the attractiveness of ODT service can be further enhanced by (a) improving the user-interface and the functionality of the service app and website, (b) giving more attention to the early-stage performance, (c) continuously updating the required fleet size based on the real-time spatio-temporal demand the system receives, (d) using medium sized vehicles rather than the current 40-ft buses, and (e) serving work-related trips independently from non-work-related trips. There are several limitations to our study, including: (a) the small sample size used in modelling, (b) the actual level of service attributes covered only the performance of the ODT service during its early stages, and (c) our analysis was limited to users who had used the ODT service at least once without including users who had only used the FRT service. In the future, we intend to improve the findings of this study by incorporating more behavioural variables, exploring different market segments, and increasing the number of observations. We also intend to develop count models for predicting the number of trips users are expected to take using the ODT service.

**ACKNOWLEDGEMENTS**

We greatly appreciate the Canadian Urban Transit Research & Innovation Consortium (CUTRIC), Toronto Metropolitan University (Formerly Ryerson University), and the Social Sciences and Humanities Research Council (SSHRC), for providing funding for this study. We are also thankful to Pantonium and the City of Belleville for providing us access to the data used in this study.

## Appendix A. Comparison between Fused and Individual Samples

The differences between the fused dataset (reduced sample) and both the operational data and the revealed preference data were verified statistically using t-test and chi-square test, respectively, and the results are presented in Tables A.1 and A.2 It should be noted that there are no significant differences between the reduced sample and the full samples at a 95% confidence level. Thus, the fused sample used in the modelling process is a representative sample for Belleville's ODT users.

**Table A.1** Statistical difference between the fused data (reduced sample) and the operational data (full sample)

| ID | Actual Level of Service Attribute | Sample Type | Average | Standard Deviation | Sample Size | t-test | $p(\|t\| \geq t_{test})$ | Significance |
|---|---|---|---|---|---|---|---|---|
| 1 | Waiting Time (min) | Full Sample | 21.69 | 19.16 | 430 | 1.87 | 0.065 | Not significant |
| | | Reduced Sample | 18.40 | 12.7 | 72 | | | |
| 2 | Assigned Trips (trips) | Full Sample | 2.62 | 5.46 | 430 | 0.358 | 0.721 | Not Significant |
| | | Reduced Sample | 2.79 | 3.37 | 72 | | | |
| 3 | Unassigned Trips (trips) | Full Sample | 6.17 | 12.97 | 430 | 1.73 | 0.088 | Not Significant |
| | | Reduced Sample | 8.39 | 9.54 | 72 | | | |

**Table A.2** Statistical difference between the fused data (reduced sample) and the revealed preference data (full sample)

| ID | Self-reported Attributes | Categories | Observed Distribution (Reduced Sample) | Expected Distribution (Full Sample) | Chi-Square Calculated | p-value | Significancy |
|---|---|---|---|---|---|---|---|
| 1 | Income (Thousand CAD) | Under 10 | 18 | 19.7 | 8.64 | 0.19 | Not Significant |
| | | 10 to 19.999 | 21 | 24.4 | | | |
| | | 20 to 29.999 | 12 | 9.4 | | | |
| | | 30 to 39.999 | 10 | 11.8 | | | |
| | | 40 to 49.999 | 5 | 3.9 | | | |
| | | 50 to 59.999 | 3 | 0.8 | | | |
| | | 60 and over | 3 | 2 | | | |
| 2 | Gender | Male | 34 | 35.2 | 4.18 | 0.12 | Not Significant |
| | | Female | 36 | 36.3 | | | |
| | | Other | 2 | 0.5 | | | |
| 3 | Marital Status | Single | 45 | 50.5 | 8.63 | 0.07 | Not Significant |
| | | Married | 9 | 11.6 | | | |
| | | Widowed | 2 | 1.7 | | | |
| | | Divorced | 5 | 3.15 | | | |
| | | Other | 11 | 5.3 | | | |
| 4 | Age | Young | 29 | 31.5 | 1.15 | 0.76 | Not Significant |
| | | Adults | 33 | 33.1 | | | |
| | | Middle-aged | 10 | 7.4 | | | |
| | | Old | 0 | 0 | | | |
| 5 | | Active Mode | 25 | 29.9 | 1.44 | 0.86 | |
| | Travel Mode | Car as a driver or passenger | 25 | 22 | | | Not Significant |
| | | FRT | 14 | 12.6 | | | |
| | | Mobility Bus Service | 3 | 2.3 | | | |
| | | Not Applicable | 5 | 4.2 | | | |
| 6 | Trip Purpose | Work-Based | 29 | 23.6 | 3.93 | 0.14 | Not Significant |
| | | Nonwork-Based | 11 | 17.9 | | | |
| | | Mixed Purposes | 32 | 30.5 | | | |
| 7 | Education level | No Formal Education | 0 | 0.5 | 10.78 | 0.06 | Not Significant |
| | | Primary School | 0 | 1.6 | | | |
| | | Secondary School | 30 | 23.1 | | | |
| | | Diploma | 24 | 17.6 | | | |
| | | Undergraduate | 9 | 15.5 | | | |
| | | Graduate | 9 | 13.7 | | | |
| 8 | Household Size | 1 person | 13 | 12.9 | 6.88 | 0.14 | Not Significant |
| | | 2 persons | 17 | 11.3 | | | |
| | | 3 persons | 14 | 11.3 | | | |
| | | 4 persons | 16 | 16.3 | | | |
| | | 5 or more persons | 12 | 20.2 | | | |

## Appendix B. The MNL Model Specification

**Table B.1 presents the final specification for the MNL utility functions.**

**Table B.1** Multinomial logit (MNL) model specification

| Parameter | Variables | | |
|---|---|---|---|
| | **ODT** | **FRT** | **Indiff.** |
| $\boldsymbol{ASC_{INDIFF}}$ | - | - | 1 |
| $\boldsymbol{ASC_{ODT}}$ | 1 | - | - |
| $\boldsymbol{\beta_{edu}}$ | - | - | $HigherEdu$ |
| $\boldsymbol{\beta_{gender}}$ | - | - | $Male$ |
| $\boldsymbol{\beta_{hhld}}$ | $Hhld_H$ | - | - |
| $\boldsymbol{\beta_{age}}$ | $MiddleAge$ | - | - |
| $\boldsymbol{\beta_{assigned_trips}}$ | $Assigned_L$ | $Assigned_H$ | - |
| $\boldsymbol{\beta_{purpose}}$ | $NonworkTrip$ | $WorkTrip$ | - |
| $\boldsymbol{\beta_{mode}}$ | $ActiveMode$ | $FixedService$ | - |
| $\boldsymbol{\beta_{in-veh}}$ | $InVeh_less$ | $InVeh_more$ | - |
| $\boldsymbol{\beta_{waiting}}$ | $Waiting_L$ | $Waiting_H$ | - |

## Appendix C. Clustering Analysis Results for the LC Model

Users can be best classified into 4 clusters based on their travel history and income level as shown in Figure C.1. The characteristics of each cluster are described in Table C.1 It can be seen that the first cluster represents low-income users who relied on other modes than public transit to satisfy their travel needs before the ODT service started. The second cluster represents medium-to- high income users who used other modes than public transit to meet their mobility needs. While medium-to- high income users who used on public transit services to satisfy their mobility needs are included in the third cluster. The fourth cluster represents low-income users who relied on public transit services to meet their travel needs.

Based on the definition of public transit captive users, described by Krizekand El-Geneidy (2007), the users belonging to the fourth cluster can be considered as ODT captive users, while those belonging to the other clusters are representing ODT non-captive users.

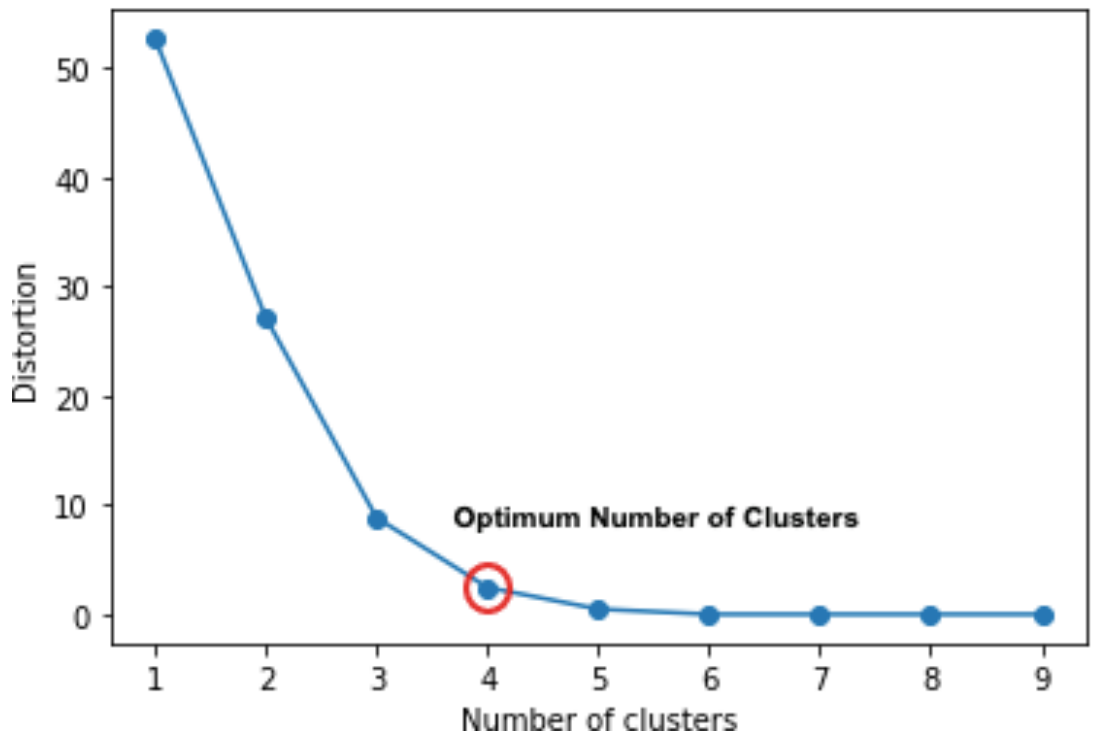

**Figure C.1** Elbow method results for users travel history and income level.

**Table C.1** Characteristics of each cluster

| Cluster | Income Level | | | Main Mode | | Counts |
|---|---|---|---|---|---|---|
| | Low Income | Medium Income | High Income | Public Transit | Other Modes | |
| 1 | 1.0 | 0.0 | 0.0 | 0.0 | 1.0 | 24 |
| 2 | 0.0 | 0.7 | 0.3 | 0.0 | 1.0 | 26 |
| 3 | 0.0 | 0.7 | 0.3 | 1.0 | 0.0 | 7 |
| 4 | 1.0 | 0.0 | 0.0 | 1.0 | 0.0 | 10 |

## Appendix D. The LC Model Specification

Table D.1 shows the class membership functions defined for the LC model, while the class-specific utility functions for alternatives are outlined in Table D.2.

**Table D.1** Class membership function specification

| Parameter | Class 1 (Captive Users) | Class 2 (Non-Captive Users) |
|---|---|---|
| $\gamma_{CAP}$ | 1 | - |
| $\gamma_{INCOME}$ | $LowIncome$ | - |
| $\gamma_{MODE}$ | $FixedService$ | - |

**Table D.2** Latent class (LC) model specification

| Parameter | Class 1 (Captive Users) | | | Class 2 (Non-Captive Users) | | |
|---|---|---|---|---|---|---|
| | ODT | FRT | Indiff. | ODT | FRT | Indiff. |
| $ASC_{ODT}^{C1}$ | 1 | - | - | - | - | - |
| $ASC_{INDIFF}^{C1}$ | - | - | 1 | - | - | - |
| $\beta_{purpose}^{C1}$ | $MixedTrip$ | $NonworkTrip$ | - | - | - | - |
| $\beta_{in-veh}^{C1}$ | $InVeh_less$ | $InVeh_more$ | - | - | - | - |
| $\beta_{waiting}^{C1}$ | $Waiting_L$ | $Waiting_H$ | - | - | - | - |
| $\beta_{unassigned_trips}^{C1}$ | $Unassigned_L$ | $Unassigned_H$ | - | - | - | - |
| $\beta_{mode}^{C1}$ | $ActiveMode$ | $FixedService$ | - | - | - | - |
| $\beta_{edu}^{C1}$ | - | - | $HigherEdu$ | - | - | - |
| $\beta_{gender}^{C1}$ | - | - | $Male$ | - | - | - |
| $ASC_{ODT}^{C2}$ | - | - | - | 1 | - | - |
| $ASC_{INDIFF}^{C2}$ | - | - | - | - | - | 1 |
| $\beta_{purpose}^{C2}$ | - | - | - | $NonworkTrip$ | $WorkTrip$ | - |
| $\boldsymbol{\beta}_{in-veh}^{C2}$ | - | - | - | $InVeh_less$ | $InVeh_more$ | - |
| $\boldsymbol{\beta}_{waiting}^{C2}$ | - | - | - | $Waiting_L$ | $Waiting_H$ | - |
| $\boldsymbol{\beta}_{assigned_trips}^{C2}$ | - | - | - | $Assigned_L$ | $Assigned_H$ | - |
| $\boldsymbol{\beta}_{edu}^{C2}$ | - | - | - | - | - | $HigherEdu$ |
| $\boldsymbol{\beta}_{gender}^{C2}$ | - | - | - | - | - | $Male$ |

**Appendix E. The ICLV Model Specification.**

Table E.1 provides the results of the principal factor analysis method. Equations (E.1) through (E.7) represent the measurement equations for the indicators presented in Table E.1. In the measurement equations, each of the latent variables Time Sensitivity and Online Services' Satisfaction is linked to its observable indicators. Table E.2 describes the specification of the structural equations for the latent variables, while Table E.3 describes the specification of the systematic utility functions for the choice alternatives.

**Table E.1** Indicators description and factor loading values.

| **ID** | **Indicator** | **Indicator Type** | **Description** | **Factor Loading** | |
|---|---|---|---|---|---|
| | | | | Time Sensitivity (TS) | Online Services' Satisfaction (OSS) |
| **1** | WAIT_IMPO | Attitude | The importance of the waiting time in influencing the user's choice of using ODT. | 0.877463 | |
| **2** | RELIA_IMPO | | The importance of the reliability in influencing the user's choice of using ODT. | 0.740253 | |
| **3** | TIME_BUS | | The importance of the in-vehicle time in influencing the user's choice of using ODT. | 0.793222 | |
| **4** | FLEXIBILITY | | The importance of the convenience and the flexibility in influencing the user's choice of using ODT. | 0.568935 | |
| **5** | APP_INTER | Perception | User's satisfaction with the ODT app interface. | | 0.894892 |
| **6** | WEB_INTER | | User's satisfaction with the ODT website interface. | | 0.845269 |
| **7** | AVAIL_SERV | | User's satisfaction with the availability of schedules/maps/fares. | | 0.475683 |

**Table E.2** Structural equation specifications

| **Parameters** | **Latent Variables** | |
|---|---|---|
| | **Time sensitivity ($TIME_SEN$)** | **Online services' satisfaction ($ON_SERV_SAT$)** |
| $\boldsymbol{\alpha}_{CONS}^{TS}$ | 1 | - |
| $\boldsymbol{\alpha}_{age}^{TS}$ | $Young$ | - |
| $\boldsymbol{\alpha}_{income}^{TS}$ | $HighIncome$ | - |
| $\boldsymbol{\alpha}_{car}^{TS}$ | $Car$ | - |
| $\boldsymbol{\alpha}_{hhld}^{TS}$ | $Hhld_L$ | - |
| $\alpha^{TS}_{gender}$ | $Male$ | - |
| $\alpha^{TS}_{marital}$ | $Single$ | - |
| $\sigma^{TS}$ | 1 | - |
| $\alpha^{OSS}_{CONS}$ | - | 1 |
| $\alpha^{OSS}_{age}$ | - | $MiddleAge$ |
| $\alpha^{OSS}_{income}$ | - | $LowIncome$ |
| $\alpha^{OSS}_{edu}$ | - | $Sec_school$ |
| $\sigma^{OSS}$ | - | 1 |

**Table E.3** Integrated Choice and Latent Variable (ICLV) model specification

| Parameter | Variables | | |
|---|---|---|---|
| | **ODT** | **FRT** | **Indiff.** |
| $ASC_{INDIFF}$ | - | - | 1 |
| $ASC_{ODT}$ | 1 | - | - |
| $\beta_{edu}$ | - | - | $HigherEdu$ |
| $\beta_{gender}$ | - | - | $Male$ |
| $\beta_{hhld}$ | $Hhld_H$ | - | - |
| $\beta_{age}$ | $MiddleAge$ | - | - |
| $\beta_{assigned_trips}$ | $Assigned_L$ | $Assigned_H$ | - |
| $\beta_{purpose}$ | $NonworkTrip$ | $WorkTrip$ | - |
| $\beta_{mode}$ | $ActiveMode$ | $FixedService$ | - |
| $\beta_{in-veh}$ | $InVeh_less$ | $InVeh_more$ | - |
| $\beta_{waiting}$ | $Waiting_L$ | $Waiting_H$ | - |
| $\beta_{TS}$ | $TIME_SEN$ | - | - |
| $\beta_{OSS}$ | $ON_SERV_SAT$ | - | - |

$$\boldsymbol{WAIT_IMPO} = \alpha_{WAIT_IMPO} + \beta_{WAIT_IMPO}\ x\ TIME_SEN + \sigma_{WAIT_IMPO}\ ; \alpha_{WAIT_IMPO} = 0, \beta_{WAIT_IMPO} = 1, \sigma_{WAIT_IMPO} = 1 \quad \textbf{(E.1)}$$

$$\boldsymbol{RELIA_IMPO} = \alpha_{RELIA_IMPO} + \beta_{RELIA_IMPO}\ x\ TIME_SEN + \sigma_{RELIA_IMPO} \quad \textbf{(E.2)}$$

$$\boldsymbol{TIME_BUS} = \alpha_{TIME_BUS} + \beta_{TIME_BUS}\ x\ TIME_SEN + \sigma_{TIME_BUS} \quad \textbf{(E.3)}$$

$$\boldsymbol{FLEXIBILITY} = \alpha_{FLEXIBILITY} + \beta_{FLEXIBILITY}\ x\ TIME_SEN + \sigma_{FLEXIBILITY} \quad \textbf{(E.4)}$$

$$\boldsymbol{APP_INTER} = \alpha_{APP_INTER} + \beta_{APP_INTER}\ x\ ON_SERV_SAT + \sigma_{APP_INTER}\ ; \alpha_{APP_INTER} = 0, \beta_{APP_INTER} = 1, \sigma_{APP_INTER} = 1 \quad \textbf{(E.5)}$$

$$\boldsymbol{WEB_INTER} = \alpha_{WEB_INTER} + \beta_{WEB_INTER}\ x\ ON_SERV_SAT + \sigma_{WEB_INTER} \quad \textbf{(E.6)}$$

$$\boldsymbol{AVAIL_SERV} = \alpha_{AVAIL_SERV} + \beta_{AVAIL_SERV}\ x\ ON_SERV_SAT + \sigma_{AVAIL_SERV} \quad \textbf{(E.7)}$$

where, $TIME_SEN$ and $ON_SERV_SAT$ are the Time Sensitivity and the Online Services' Satisfaction latent variables, respectively. $\boldsymbol{\sigma}_{WAIT_IMPO}$ , $\boldsymbol{\sigma}_{RELIA_IMPO}$ , $\boldsymbol{\sigma}_{TIME_BUS}$ , $\boldsymbol{\sigma}_{FLEXIBILITY}$ , $\boldsymbol{\sigma}_{APP_INTER}$ , $\boldsymbol{\sigma}_{WEB_INTER}$, and $\boldsymbol{\sigma}_{AVAIL_SERV}$ are scale parameters to be estimated. It is worth mentioning that the first measurement equation of the Time Sensitivity and the Online Services' Satisfaction latent variables (WAIT_IMPO and APP_INTER equations, respectively) were normalized by setting their constants to (0.00) and their coefficients and scale parameters to (1.00) for identification purposes (Ababio-Donkor et al., 2020; Alizadeh et al., 2019; Bierlaire, 2018).

## Appendix F. Model Estimates

Tables F.1 and F.2 provide the estimation results for the MNL and ICLV models, while Table F.3 presents the estimation results for the LC model.

**Table F.1** Estimation results of the measurement equations of the ICLV model

| Indicator | Parameter | Estimate | Rob. t-test |
|---|---|---|---|
| WAIT_IMPO | $\alpha_{\text{WAIT_IMPO}}$ | 0.000 | – |
| | $\beta_{\text{WAIT_IMPO}}$ | 1.000 | – |
| | $\sigma_{\text{WAIT_IMPO}}$ | 1.000 | – |
| RELIA_IMPO | $\alpha_{\text{RELIA_IMPO}}$ | 1.100 | 1.93* |
| | $\beta_{\text{RELIA_IMPO}}$ | 0.874 | 1.64* |
| | $\sigma_{\text{RELIA_IMPO}}$ | 1.530 | 3.26 |
| TIME_BUS | $\alpha_{\text{TIME_BUS}}$ | -0.140 | -0.41** |
| | $\beta_{\text{TIME_BUS}}$ | 0.674 | 2.04 |
| | $\sigma_{\text{TIME_BUS}}$ | 1.230 | 6.38 |
| FLEXIBILITY | $\alpha_{\text{FLEXIBILITY}}$ | 1.080 | 2.05 |
| | $\beta_{\text{FLEXIBILITY}}$ | 0.051 | 0.12** |
| | $\sigma_{\text{FLEXIBILITY}}$ | 1.31 | 5.35 |
| APP_INTER | $\alpha_{\text{APP_INTER}}$ | 0.000 | – |
| | $\beta_{\text{APP_INTER}}$ | 1.000 | – |
| | $\sigma_{\text{APP_INTER}}$ | 1.000 | – |
| WEB_INTER | $\alpha_{\text{WEB_INTER}}$ | 0.097 | 1.15** |
| | $\beta_{\text{WEB_INTER}}$ | 0.843 | 7.48 |
| | $\sigma_{\text{WEB_INTER}}$ | 0.215 | 4.96 |
| AVAIL_SERV | $\alpha_{\text{AVAIL_SERV}}$ | 0.351 | 2.85 |
| | $\beta_{\text{AVAIL_SERV}}$ | 0.599 | 4.68 |
| | $\sigma_{\text{AVAIL_SERV}}$ | 0.732 | 7.57 |

* Not statistically significant at 95% confidence level

** Not statistically significant at 90% confidence level

**Table F.2** Estimation results of the choice and latent variable models

| Parameter | MNL | | ICLV | |
|---|---|---|---|---|
| | Estimate | Rob. t-test | Estimate | Rob. t-test |
| **Choice Model** | | | | |
| $ASC_{INDIFF}$ | -1.89 | -2.60 | -1.89 | -2.64 |
| $ASC_{ODT}$ | -2.61 | -2.71 | -3.66 | -2.68 |
| $\beta_{assigned_trips}$ | 0.972 | 1.54** | 0.927 | 1.43** |
| $\beta_{purpose}$ | 1.09 | 2.40 | 1.12 | 2.25 |
| $\beta_{mode}$ | 1.03 | 2.31 | 1.17 | 2.46 |
| $\beta_{in-veh}$ | 1.35 | 3.23 | 1.45 | 3.18 |
| $\beta_{waiting}$ | 1.07 | 2.12 | 1.14 | 2.32 |
| $\beta_{hhld}$ | 2.11 | 2.43 | 2.46 | 2.43 |
| $\beta_{age}$ | 1.91 | 1.82* | 1.92 | 1.79* |
| $\beta_{edu}$ | 1.55 | 2.07 | 1.49 | 2.03 |
| $\beta_{gender}$ | 2.17 | 2.70 | 2.38 | 2.82 |
| $\beta_{TS}$ | - | - | 0.668 | 1.29** |
| $\beta_{OSS}$ | - | - | 0.385 | 1.12** |
| **Latent Variable Model (Structural Equations)** | | | | |
| $\alpha^{TS}_{CONS}$ | - | - | 0.334 | 0.92** |
| $\alpha^{TS}_{age}$ | - | - | 1.080 | 2.29 |
| $\alpha^{TS}_{income}$ | - | - | 0.687 | 2.14 |
| $\alpha^{TS}_{car}$ | - | - | -0.620 | -1.21** |
| $\alpha^{TS}_{hhld}$ | - | - | 0.724 | 2.14 |
| $\alpha^{TS}_{gender}$ | - | - | 1.110 | 2.77 |
| $\alpha^{TS}_{marital}$ | - | - | -1.040 | -2.31 |
| $\sigma^{TS}$ | - | - | -0.113 | -0.48** |
| $\alpha^{OSS}_{CONS}$ | - | - | -0.858 | -3.85 |
| $\alpha^{OSS}_{age}$ | - | - | 1.290 | 3.34 |
| $\alpha^{OSS}_{income}$ | - | - | 1.360 | 4.94 |
| $\alpha^{OSS}_{edu}$ | - | - | 0.640 | 3.62 |
| $\sigma^{OSS}$ | - | - | 1.300 | 7.69 |
| **Performance Indicators** | | | | |
| Number of observations | 72 | | 72 | |
| Number of parameters | 11 | | 41 | |
| Initial log-likelihood | -79.10 | | -882.69 | |
| Final log-likelihood | -47.49 | | -577.93 | |
| Rho-square-bar | 0.26 | | 0.299 | |

* Not statistically significant at 95% confidence level

** Not statistically significant at 90% confidence level

**Table F.3** Estimation results of the choice model and class membership functions for the LC model

| **Parameters** | **Estimate** | **Rob. t-test** |
|---|---|---|
| **Choice Model** | | |
| $ASC^{C1}_{ODT}$ | -11.600 | -11.30 |
| $ASC^{C1}_{INDIFF}$ | -2.430 | -2.01 |
| $\beta^{C1}_{purpose}$ | 0.893 | 1.54** |
| $\beta^{C1}_{in-veh}$ | 1.440 | 2.35 |
| $\beta^{C1}_{waiting}$ | 0.386 | 0.73** |
| $\beta^{C1}_{unassigned_trips}$ | 10.300 | 9.88 |
| $\beta^{C1}_{mode}$ | 1.450 | 2.14 |
| $\beta^{C1}_{edu}$ | 2.280 | 1.87* |
| $\beta^{C1}_{gender}$ | 1.010 | 0.77** |
| $\beta^{C1}_{TS}$ | - | - |
| $\beta^{C1}_{OSS}$ | - | - |
| $ASC^{C2}_{ODT}$ | 1.700 | 1.91* |
| $ASC^{C2}_{INDIFF}$ | -9.100 | -14.50 |
| $\beta^{C2}_{purpose}$ | 10.900 | 4.26 |
| $\beta^{C2}_{in-veh}$ | 6.560 | 3.21 |
| $\beta^{C2}_{waiting}$ | 3.000 | 2.14 |
| $\beta^{C2}_{assigned_trips}$ | 2.330 | 2.12 |
| $\beta^{C2}_{edu}$ | 13.300 | 10.20 |
| $\beta^{C2}_{gender}$ | 17.700 | 7.93 |
| $\beta^{C2}_{TS}$ | - | - |
| $\beta_{OSS}^{C2}$ | - | - |
| **Class Membership Functions** | | |
| $\gamma_{CAP}$ | -10.600 | -21.30 |
| $\gamma_{INCOME}$ | 24.100 | 20.70 |
| $\gamma_{MODE}$ | 21.600 | 32.50 |
| **Performance Indicators** | | |
| Number of observations | 72 | |
| Number of parameters | 20 | |
| Initial log-likelihood | -79.10 | |
| Final log-likelihood | -36.47 | |
| Rho-square-bar | 0.286 | |

* Not statistically significant at 95% confidence level

** Not statistically significant at 90% confidence level